\documentclass[a4paper,10pt,12pt]{article}
\usepackage{amsmath, amssymb, latexsym, amscd, amsthm,amsfonts,amstext}
\usepackage[mathscr]{eucal}
\usepackage{graphicx}
\usepackage{subfig}

\numberwithin{equation}{section}
\input{tcilatex}
\begin{document}

\title{Carleman estimates for the regularization of ill-posed Cauchy
problems }
\author{Michael V. Klibanov \\
\\
Department of Mathematics \& Statistics, University of North Carolina at
Charlotte,\\
Charlotte, NC, USA. Email: \texttt{\ mklibanv@uncc.edu}}
\date{}
\maketitle

\begin{abstract}
This is a survey, which is a continuation of the previous survey of the
author about applications of Carleman estimates to Inverse Problems, J.
Inverse and Ill-Posed Problems, 21, 477-560, 2013. It is shown here that
Tikhonov functionals for some ill-posed Cauchy problems for linear PDEs can
be generated by unbounded linear operators of those PDEs. These are those
operators for which Carleman estimates are valid, e.g. elliptic, parabolic
and hyperbolic operators of the second order. Convergence rates of
minimizers are established using Carleman estimates. Generalizations to
nonlinear inverse problems, such as problems of reconstructions of obstacles
and coefficient inverse problems are discussed as well.
\end{abstract}

\textbf{Keywords}: Survey, Carleman estimates, Ill-Posed Cauchy problems,
convergence rates.

\textbf{AMS classification codes:} 65N15, 65N30, 35J25.

\graphicspath{{Figures/}}

\graphicspath{{FIGURES/}
{Figures/}
{FiguresJ/newfigures/}
{pics/}}

\section{Introduction}

\label{sec:1}

This work is a survey, which is the continuation of the recent survey \cite%
{Ksurvey} of the author about applications of the method of Carleman
estimates to inverse problems. Let $\Psi \subset \mathbb{R}^{n}$ be a
bounded domain and $A$ be a linear Partial\ Differential Operator (PDO) of
the second order acting in $\Psi $. Likewise, assume that this operator
admits a Carleman estimate. In fact, the class of such operators is quite
broad.\ Indeed, currently Carleman estimates are \ derived for three main
classes of PDOs of the second order: elliptic, parabolic and hyperbolic
ones, see, e.g. books of Beilina and Klibanov \cite{BK}, Isakov \cite{Is},
Klibanov and Timonov \cite{KT} and Lavrentiev, Romanov and Shishatskii \cite%
{LRS} as well as the paper of Triggiani and Yao \cite{Trig}. Therefore,
results of this paper are quite general ones. Consider the Partial\
Differential Equation (PDE) $Au=f$ and an ill-posed Cauchy problem for it.
Suppose that the Tikhonov functional for the solution of this problem is
generated by the operator $A$.\emph{\ }The current paper provides the
positive answer for the following question: \emph{Can the solution of this
Cauchy problem be approximated via the minimization of this functional?}

Typically the regularization term is presented in the Tikhonov functional in
a norm, which is stronger than the norm of the original space. Hence, we
consider in our setting\emph{\ }the domain of $A$ as $D\left( A\right)
=H^{2}\left( \Psi \right) \subset L_{2}\left( \Psi \right) $ and $%
A:H^{2}\left( \Psi \right) \rightarrow L_{2}\left( \Psi \right) .$ Thus, in
this specific context $H^{2}\left( \Psi \right) $ is a linear set in $%
L_{2}\left( \Psi \right) $ and $\overline{H}^{2}\left( \Psi \right)
=L_{2}\left( \Psi \right) ,$ where the closure is taken in the norm of $%
L_{2}\left( \Psi \right) .$ Thus, we consider $A$ as an unbounded operator.
As to the regularization theory for linear ill-posed problems with bounded
linear operators, we refer to, e.g. the book of Ivanov, Vasin and Tanana 
\cite{Iv}.

For PDOs of the second order with above operators $A$, we present here a
universal method of both constructions of Tikhonov functionals for solutions
of ill-posed Cauchy problems for corresponding PDEs and estimating
convergence rates of minimizers. First, we present our universal approach in
which the operator $A$ of the original PDE generates the Tikhonov
functional. Next, we specify our method for four (4) main classes of
ill-posed Cauchy problems: Cauchy problems for elliptic PDEs, problems for
hyperbolic and parabolic PDEs with the lateral Cauchy data and the initial
boundary value problem for the parabolic PDE with the reversed time. In
addition, we briefly outlinw in subsections 8.2, 8.3 extensions of this
method to two important nonlinear inverse problems: inverse obstacle
problems and coefficient inverse problems.

Unlike the current paper, the survey \cite{Ksurvey} was focused on the
method, which was first proposed by Bukhgeim and Klibanov \cite%
{BukhK,Bukh1,Klib1} for the (papers \cite{Bukh1,Klib1} contain first full
proofs). The method of \cite{BukhK} is based on Carleman estimates. The
specific topic of the current paper was only briefly mentioned on pages
496-498 of \cite{Ksurvey}. The method of \cite{BukhK,Bukh1,Klib1} was
originally designed for proofs of uniqueness theorems for Coefficient
Inverse Problems (CIPs) with single measurement data, see, e.g. some follow
up works of Bukhgeim \cite{Bukh}, Klibanov \cite{Klib3,K92}, Klibanov and
Timonov \cite{KT}, a survey of Yamamoto \cite{Y}, as well as sections 1.10
and 1.11 of the book of Beilina and Klibanov \cite{BK}. Later, this idea was
extended to globally convergent numerical methods for CIPs, see works of the
author with coauthors \cite{BKconv,Klib97,Kpar,KT,KNT}, the paper of
Baudouin, de Buhan and Ervedoza \cite{Baud} and subsection 8.3.

The role of Carleman estimates in our universal regularization method is
that they provide convergence rates of minimizers of those Tikhonov
functionals. The true reason why Carleman estimates are so helpful here is
that they provide H\"{o}lder stability estimates in certain subdomains for
those Cauchy problems for elliptic and parabolic PDEs, see, e.g. \cite%
{Is,KT,Ksurvey,LRS}. In the hyperbolic case the Carleman estimate provides
even the stronger Lipschitz stability estimate in the whole domain, see this
section below and section 6. It turns out that the H\"{o}lder stability
estimate in a subdomain is a certain analog of the estimate of the modulus
of the continuity of the inverse operator. On the other hand, it is one of
classical results of the theory of ill-posed problems that an estimate of
the modulus of the continuity on a compact set of the inverse operator
provides the rate of convergence of minimizers of the Tikhonov functional,
see, e.g. books of Beilina and Klibanov \cite{BK},\ Engl, Hanke and Neubauer 
\cite{EHN}, Kabanikhin \cite{Kab}, Lavrentiev, Romanov and Shishatskii \cite%
{LRS} and Tikhonov, Goncharsky, Stepanov and Yagola \cite{Tikh}.

The first Tikhonov functionals for ill-posed Cauchy problems for PDEs, which
were generated by operators of those PDEs, were constructed in the \emph{%
pioneering} work of Lattes and Lions \cite{LL}. Lattes and Lions have called
their approach the \textquotedblleft Quasi-Reversibility Method" (QRM).
Their book contains examples of quite many ill-posed Cauchy problems. They
have presented two versions of the QRM. In the first version, additional
terms with regularization parameters in them were introduced in those PDEs.\
In the second version, strong formulations of those Cauchy problems were
considered first, in which the fourth order operator $A^{\ast }A$ is
involved. In the latter case, for elliptic and parabolic PDEs, weak
variational formulations of equations with $A^{\ast }A$ were considered
next. In the elliptic case, that variational formulation was equivalent with
the minimization of a Tikhonov functional generated by the operator $A$. In
the parabolic case, the original strong formulation led to an unnecessary
term $\left( u\left( T\right) ,v\left( T\right) \right) $ in the variational
form, see the formula (3.6) on page 324 of \cite{LL}.\ Also, certain cut-off
functions were used in \cite{LL}, which is unnecessary. Convergence theorems
were proved in \cite{LL}. Convergence rates were not established in \cite{LL}
and Carleman estimates were not used.

First applications of the tool of Carleman estimates to this topic were done
in papers of Klibanov and Santosa \cite{KS} and Klibanov and Malinsky \cite%
{KM}. As a result, first convergence rates of minimizers of those Tikhonov
functionals were established in these references. Both these works have
considered the variational form of the Tikhonov functional. In \cite{KS} the
Cauchy problem for the Laplace operator was considered, see the paper of
Cao, Klibanov and Pereverzev \cite{Cao} for a continuation of \cite{KS}.
Since in the elliptic case the H\"{o}lder stability estimate can be proved
by the Carleman estimate only in a subdomain, then H\"{o}lder-like
convergence rates of minimizers in \cite{Cao,KS} were established only in
subdomains; also see sections 2-5 below.

The paper \cite{KM} is the first one where the Lipschitz stability estimate\
in the entire time cylinder was proved for the hyperbolic equation with the
lateral Cauchy data, using a Carleman estimate (see Theorem 6.1 in section
6). Given the Carleman estimate, the Lipschitz stability estimate became
possible basically because the hyperbolic equation can be stably solved in
both directions of time: positive and negative. The Lipschitz stability
estimate, in turn allows to establish Lipschitz-like convergence rates of
minimizers of the corresponding Tikhonov functional in the entire time
cylinder, see \cite{KM} and section 6. There were several follow up works,
which explored some modifications of the idea of \cite{KM} to prove the
Lipschitz stability for the hyperbolic case. More precisely, those were
works of Kazemi and Klibanov \cite{Kaz}, Klibanov and Timonov \cite{KT},
Isakov \cite{Is}, Clason and Klibanov \cite{ClK}, Klibanov, Kuzhuget,
Kabanikhin and Nechaev \cite{KKKN} and the survey \cite{Ksurvey}. While all
these publications are about the case of the Euclidian geometry, the more
general case of the Riemannian geometry was considered by Triggiani and Yao 
\cite{Trig}. Lasiecka, Triggiani and Zhang \cite{LT3,LT4} have extended this
technique to the case of the Schr\"{o}dinger equation.

It is shown in section 7 that the original technique of \cite{KM} allows one
to obtain the Lipschitz stability estimate, to construct the Tikhonov
functional and to obtain the Lipschitz-like convergence rate of its
minimizers for the problem of determining an initial condition of a
hyperbolic PDE from boundary measurements.\ This problem is called nowadays
\textquotedblleft the problem of thermoacoustic tomography"; also, see more
details in the paper of the author \cite{Kltherm}.

Based on the ideas of \cite{KS,KM}, a universal regularization method for
ill-posed Cauchy problems was developed in the book \cite{KT}. Later, it was
briefly discussed in \cite{Ksurvey}. This method works for those PDEs, for
which Carleman estimates are valid. On the first step of this approach such
a Tikhonov functional is constructed which is generated by the unbounded
operator of the corresponding PDE. On the second step the convergence rate
of minimizers of that functional is established using a corresponding
Carleman estimate. Unlike \cite{LL}, cut-off functions are not used in this
method. We now refer to papers of the author with coauthors \cite%
{Cao,ClK,KR,Klib2006,KKKN,Kltherm}, which have explored this method.

As to the numerical implementation of our regularization method, it can work
either with the finite difference, or with the finite element version, or
with the spline formulation of that Tikhonov functional. The case of finite
differences was implemented in \cite{KS,LL}, in the paper of Klibanov and
Rakesh \cite{KR} and in the paper of Klibanov, Kuzhuget, Kabanikhin and
Nechaev \cite{KKKN}. As to the finite elements, see, e.g. Bourgeois \cite{B1}
and more details in section 8.\ Cao, Klibanov and Pereverzev \cite{Cao} and
Clason and Klibanov \cite{ClK} have used cubic $B-$splines in numerical
studies. In particular, in papers \cite{ClK,KKKN} the problem of
thermoacoustic tomography was numerically studied. Numerical testing has
always shown a very good performance.

There are a large number of publications discussing solutions of ill-posed
Cauchy problems for PDEs. Since the focus of this paper is on solving these
problems via minimizations of Tikhonov functionals generated by
corresponding PDE operators, the author refers here only to a few of such
works. Other references can be found in, e.g. \cite{Kab}.

In a number of works of Kabanikhin with coauthors, which were summarized in
the book \cite{Kab}, a variety of ill-posed Cauchy problems was solved via
minimizations of various forms of the Tikhonov functional. Naturally, that
form depends on a specific PDE; also, see, e.g. Kabanikhin and Shishlenin 
\cite{KabSh} and Karchevsky \cite{Karch}. In \cite{Kab,KabKar,KabSh} the
operator $A$ generating the Tikhonov functional is the one, which
establishes the correspondence between the sought for boundary data $q$ on
an inaccessible part of the boundary and an extra boundary condition $f$ in
the given Cauchy data on the accessible part of the boundary, $Aq=f$. Thus, $%
A$ is a linear bounded operator in this case. Hao \cite{Hao1} as well as Hao
and Lesnic \cite{HaoLes} have \ published a similar approach for a parabolic
equation with the lateral Cauchy data and for the Cauchy problem for the
Laplace equation. Yagola, Leonov and Titarenko \cite{Yag} have \ studied the
heat equation with the reversed time and the Cauchy problem for the Laplace
equation using them as some specific examples of the application of the
general theory of Tikhonov functionals for linear bounded operators.

Kozlov, Maz'ya and Fomin \cite{Kozlov1} have proposed, for the first time,
an alternated iterative method for the Cauchy problem for elliptic
equations.\ This method has gained a lot of popularity since then. In this
regard, we also mention the work of Avdonin, Kozlov and Maxwell \cite%
{Kozlov2} for a \ nonlinear elliptic equation and the work of Berntsson,
Kozlov, Mpinganzima and Turesson \cite{Kozlov3} for the Helmholtz equation.
Andrieux, Baranger and Ben Abda \cite{A} have improved in some sense the
algorithm of \cite{Kozlov1}. Lie, Hie and Zou in their elegant work \cite%
{Zou} have constructed a version of the Adaptive Finite Element Method
(adaptivity) for the Cauchy problem for an elliptic equation. As to the
adaptivity for CIPs, see, e.g. \cite{BK,BKadap} for studies of experimental
data.

Bakushinsky and Gonchasky in their book \cite{Bak} have constructed
regularizing algorithms with operators $A^{\ast }A$ in them for solving
ill-posed problems in Hilbert spaces for equations with unbounded abstract
operators $A$ acting in Hilbert spaces. Also, Bakushinsky \cite{Bak1} has
originated the method of solving ill-posed Cauchy problems for abstract
operator equations in Banah spaces using finite differences with respect to
one variable. In this approach the grid step size is linked with the level
of error in the data, which is natural for ill-posed problems. The interest
to this idea was recently renewed, see, e.g. two papers of Bakushinsky,
Kokurin and Kokurin \cite{Bak2,Bak3} and references cited there. They have
shown that their procedures are stable and estimated rates of convergence.\
Numerical results are presented in \cite{Bak2,Bak3}.

Eld\'{e}n \cite{Elden1} and then Eld\'{e}n, Berntsson and Regi\'{n}ska \cite%
{Elden2} have proposed to solve the 1-d parabolic equation with the lateral
Cauchy data on one edge of a spatial interval via considering the Fourier
transform with respect to time first and then solving the Cauchy problem for
the resulting PDE, using some regularization. The H\"{o}lder stability
estimate was obtained in \cite{Elden1}. Note that the technique of \cite%
{Elden2} works for spatially dependent coefficients. Furthermore, an
interesting numerical example of \cite{Elden2} demonstrates a successful
performance of this method on some experimental data.

The structure of this survey is the following. In section 2 we describe our
universal regularization method for a generic linear PDO $A$ of the second
order.\ Next, in sections 3-6 we illustrate how this method works for four
main classes of ill-posed Cauchy problems, which are mentioned in the first
paragraph of this section. In section 7 we discuss the problem of
thermoacoustic tomography. Finally, in section 8 we briefly outline other
results regarding the topic of this survey. Thus, an interested reader would
read those results in original publications in detail. In particular, we
describe in section 8 two classes of nonlinear inverse problems, to which
some modifications of that universal regularization method are applicable.\
Those classes are: inverse obstacle problems and CIPs. All functions
considered below are real valued ones.

\section{The Universal Regularization Method}

\label{sec:2}

\subsection{The Carleman estimate}

\label{sec:2.1}

We now introduce the notion of the pointwise Carleman estimate for a general
Partial Differential Operator of the second order. Let $\Omega \subset 
\mathbb{R}^{n}$ be a bounded domain with a piecewise smooth boundary $%
\partial \Omega $. Let the function $\xi \in C^{2}\left( \overline{\Omega }%
\right) $ and $\left\vert \nabla \xi \right\vert \neq 0$ in $\overline{%
\Omega }.$ In a Carleman estimate, an important role is played by level
surface of the Carleman Weight Function (CWF). For a number $c\geq 0$ and a
function $\xi \left( x\right) $ defined in $\Omega $ denote 
\begin{equation}
\xi _{c}=\left\{ x\in \overline{\Omega }:\xi \left( x\right) =c\right\}
,\Omega _{c}=\left\{ x\in \Omega :\xi \left( x\right) >c\right\} .
\label{2.0}
\end{equation}%
We assume that $\Omega _{c}\neq \varnothing .$ Let $\Gamma _{c}\subseteq
\partial \Omega ,\Gamma _{c}\in C^{1}$ be a part of the boundary $\partial
\Omega $ defined as $\Gamma _{c}=\left\{ x\in \partial \Omega :\xi \left(
x\right) >c\right\} .$ We assume that $\Gamma _{c}\neq \varnothing .$ Then
the boundary of the domain $\Omega _{c}$ consists of two parts,%
\begin{equation}
\partial \Omega _{c}=\partial _{1}\Omega _{c}\cup \partial _{2}\Omega
_{c},\partial _{1}\Omega _{c}=\xi _{c},\partial _{2}\Omega _{c}=\Gamma _{c}.
\label{2.1}
\end{equation}%
Let $\lambda >1$ be a large parameter. Consider the function $\varphi
_{\lambda }\left( x\right) ,$%
\begin{equation}
\varphi _{\lambda }\left( x\right) =\exp \left( \lambda \xi \left( x\right)
\right) .  \label{2.2}
\end{equation}%
It follows from (\ref{2.1}), (\ref{2.2}) that 
\begin{equation}
\min_{\overline{\Omega }_{c}}\varphi _{\lambda }\left( x\right) =\varphi
_{\lambda }\left( x\right) \mid _{\xi _{c}}=e^{\lambda c}.  \label{2.3}
\end{equation}%
Let $A\left( x,D\right) $ be a linear PDO of the second order in $\Omega $
with its principal part $A_{0}\left( x,D\right) .$ We assume below that $%
\Gamma _{c}$ is a non-characteristic hypersurface for the operator $A_{0}$,
where 
\begin{eqnarray}
A\left( x,D\right) u &=&\sum\limits_{\left\vert \alpha \right\vert \leq
2}a_{\alpha }\left( x\right) D^{\alpha }u,\text{ }A_{0}\left( x,D\right)
u=\sum\limits_{\left\vert \alpha \right\vert =2}a_{\alpha }\left( x\right)
D^{\alpha }u.  \label{2.100} \\
a_{\alpha } &\in &C^{1}\left( \overline{\Omega }\right) \text{ for }%
\left\vert \alpha \right\vert =2\text{, }a_{\alpha }\in C\left( \overline{%
\Omega }\right) \text{ for }\left\vert \alpha \right\vert =0,1.
\label{2.101}
\end{eqnarray}

\textbf{Definition 2.1}. \emph{Let }$\Omega _{c}\neq \varnothing .$ \emph{We
say that the operator }$A_{0}\left( x,D\right) $\emph{\ admits pointwise\
Carleman estimate in the domain }$\Omega _{c}$\emph{\ with the Carleman
Weight Function (CWF) }$\varphi _{\lambda }\left( x\right) $\emph{\ if there
exist constants }$\lambda _{0}\left( \Omega _{c},A_{0}\right) >1,C_{1}\left(
\Omega _{c},A_{0}\right) >0$\emph{\ depending only on the domain }$\Omega $%
\emph{\ and the operator }$A_{0}$\emph{, such that the following a priori
estimate holds}%
\begin{eqnarray}
\left( A_{0}u\right) ^{2}\varphi _{\lambda }^{2}\left( x\right) &\geq
&C_{1}\lambda \left( \nabla u\right) ^{2}\varphi _{\lambda }^{2}\left(
x\right) +C_{1}\lambda ^{3}u^{2}\varphi _{\lambda }^{2}\left( x\right) +%
\func{div}U,  \label{2.6} \\
\forall \lambda &\geq &\lambda _{0},\forall u\in C^{2}\left( \overline{%
\Omega }\right) ,\forall x\in \Omega _{c}.  \label{2.7}
\end{eqnarray}%
\emph{In (\ref{2.6}) vector function }$U\left( x\right) $ \emph{satisfies
the following estimate} 
\begin{equation}
\left\vert U\left( x\right) \right\vert \leq C_{1}\lambda ^{3}\left[ \left(
\nabla u\right) ^{2}+u^{2}\right] \varphi _{\lambda }^{2}\left( x\right)
,\forall x\in \Omega _{c}.  \label{2.8}
\end{equation}

\textbf{Lemma 2.1}.\emph{\ Let conditions (\ref{2.100}), (\ref{2.101}) hold.
Suppose that the pointwise Carleman estimate (\ref{2.6})-(\ref{2.8}) is
valid for the principal part }$A_{0}\left( x,D\right) $\emph{\ of the
operator }$A\left( x,D\right) .$\emph{\ Then this estimate is also valid \
for the operator }$A\left( x,D\right) $\emph{, although with a different
constant }$\lambda _{0}.$ \emph{In other words, the Carleman estimate
depends only on the principal part of the operator. }

\textbf{Proof.} This lemma is elementary and well known. We have 
\begin{equation}
\left( Au\right) ^{2}\varphi _{\lambda }^{2}\left( x\right) \geq \left(
A_{0}u\right) ^{2}\varphi _{\lambda }^{2}\left( x\right) -M\left[ \left(
\nabla u\right) ^{2}+u^{2}\right] \varphi _{\lambda }^{2}\left( x\right)
,\forall x\in \Omega _{c},  \label{2.9}
\end{equation}%
where $M>0$ is a constant depending only on the maximum of norms $\left\Vert
a_{\alpha }\right\Vert _{C\left( \overline{\Omega }\right) },\left\vert
\alpha \right\vert =0,1.$ Comparing (\ref{2.9}) with (\ref{2.6}) and taking $%
\lambda $ sufficiently large, we obtain such an analog of (\ref{2.6}) in
which $A_{0}u$ is replaced with $Au$. $\square $

\subsection{H\"{o}lder stability}

\label{sec:2.2}

\bigskip Consider the following Cauchy problem for the differential
inequality%
\begin{eqnarray}
\left\vert A_{0}u\right\vert  &\leq &B\left( \left\vert \nabla u\right\vert
+\left\vert u\right\vert +\left\vert f\right\vert \right) \text{ in }\Omega
_{c},u\in H^{2}\left( \Omega _{c}\right) ,  \label{2.10} \\
u &\mid &_{\Gamma _{c}}=g_{0}\left( x\right) ,\partial _{n}u\mid _{\Gamma
_{c}}=g_{1}\left( x\right) ,  \label{2.11}
\end{eqnarray}%
where $B=const.>0$ and the function $f\in L_{2}\left( \Omega _{c}\right) $.
Since $\Gamma _{c}$ is a non-characteristic hypersurface for the operator $%
A_{0},$ then functions $g_{0},g_{1}$ in (\ref{2.11}) are the Cauchy data for
the function $u$. Obviously, equation $Au=f$ with the boundary data (\ref%
{2.11}) can be reduced to the problem (\ref{2.10}), (\ref{2.11}). We now
estimate the function $u$ via functions $f,g_{0},g_{1}.$ Such estimates were
derived for parabolic, elliptic and hyperbolic operators in Chapter 4 of the
book of Lavrent'ev, Romanov and Shishatskii \cite{LRS}, in section 2.3 of 
\cite{KT} and in \cite{Ksurvey}.

\textbf{Theorem 2.1} (H\"{o}lder stability estimate). \emph{Assume that
conditions (\ref{2.100}), (\ref{2.101}) hold and that the Carleman estimate
of Definition 2.1 is valid.\ Suppose that there exists a sufficiently small
number }$\varepsilon >0$\emph{\ such that }$\Omega _{c+3\varepsilon }\neq
\varnothing $ \emph{and} $\Gamma _{c+3\varepsilon }\neq \varnothing .$\ I%
\emph{n addition, assume that }$g_{0}\in H^{1}\left( \Gamma _{c}\right)
,g_{1}\in L_{2}\left( \Gamma _{c}\right) ,f\in L_{2}\left( \Omega
_{c}\right) .$\emph{\ Let }$m=\max_{\overline{\Omega }_{c}}\xi \left(
x\right) $\emph{\ and }$\beta =\left( 2\varepsilon \right) /\left(
3m+2\varepsilon \right) \in \left( 0,1\right) .$ \emph{Assume that the
function }$u\in H^{2}\left( \Omega _{c}\right) $\emph{\ satisfies conditions
(\ref{2.10}), (\ref{2.11}). Then there exists a sufficiently small number }$%
\delta _{0}=\delta _{0}\left( \varepsilon ,m,B,A_{0},\Omega _{c}\right) \in
\left( 0,1\right) $\emph{\ and a constant }$C_{2}=C_{2}\left( \varepsilon
,m,B,A_{0},\Omega _{c}\right) >0$\emph{\ such that if for }$\delta \in
\left( 0,\delta _{0}\right) $\emph{\ }%
\begin{equation}
\left\Vert f\right\Vert _{L_{2}\left( \Omega _{c}\right) },\left\Vert
g_{0}\right\Vert _{H^{1}\left( \Gamma _{c}\right) },\left\Vert
g_{1}\right\Vert _{L_{2}\left( \Gamma _{c}\right) }\leq \delta ,
\label{2.11_1}
\end{equation}%
\emph{then the following H\"{o}lder stability estimate is valid}%
\begin{equation}
\left\Vert u\right\Vert _{H^{1}\left( \Omega _{c+3\varepsilon }\right) }\leq
C_{2}\left( 1+\left\Vert u\right\Vert _{H^{1}\left( \Omega _{c}\right)
}\right) \delta ^{\beta },\forall \delta \in \left( 0,\delta _{0}\right) .
\label{2.11_2}
\end{equation}

\textbf{Proof}. In this proof, $C_{1}=C_{1}\left( \Omega _{c},A_{0}\right) $
and $C_{2}=C_{2}\left( \varepsilon ,m,B,A_{0},\Omega _{c}\right) $ denote
different positive constants depending on listed parameters. Since $\Omega
_{c+3\varepsilon }\neq \varnothing $ and $\Omega _{c+3\varepsilon }\subset
\Omega _{c+2\varepsilon }\subset \Omega _{c+\varepsilon }\subset \Omega _{c},
$ then $\Omega _{c+2\varepsilon },\Omega _{c+\varepsilon },\Omega _{c}\neq
\varnothing .$ Obviously $\overline{\Omega }_{c+3\varepsilon }\cap \partial
\Omega =\Gamma _{c+3\varepsilon }\subset \Gamma _{c}.$ Recall that $\Gamma
_{c+3\varepsilon }\neq \varnothing .$ Choose the function $\chi \left(
x\right) $ such that%
\begin{equation}
\chi \in C^{2}\left( \overline{\Omega }_{c}\right) ,\chi \left( x\right)
=\left\{ 
\begin{array}{c}
1,x\in \Omega _{c+2\varepsilon }, \\ 
0,x\in \Omega _{c}\diagdown \Omega _{c+\varepsilon }, \\ 
\in \left[ 0,1\right] ,x\in \Omega _{c+\varepsilon }\diagdown \Omega
_{c+2\varepsilon }.%
\end{array}%
\right.   \label{2.11_3}
\end{equation}%
The existence of such functions is well known from the Real Analysis course.
First, let $u\in C^{2}\left( \overline{\Omega }\right) .$ Consider the
function $v=\chi u.$ Representing $u=\chi u+\left( 1-\chi \right) u=v+\left(
1-\chi \right) u$ and using (\ref{2.10}), (\ref{2.11}) and (\ref{2.11_3}),
we obtain%
\begin{equation}
\left\vert A_{0}v\right\vert \leq C_{2}\left[ \left\vert \nabla v\right\vert
+\left\vert v\right\vert +\sum\limits_{\left\vert \alpha \right\vert \leq
2}\left\vert D^{\alpha }\left( 1-\chi \right) u\right\vert +\left\vert
f\right\vert \right] ,\forall x\in \Omega _{c},  \label{2.11_4}
\end{equation}%
\begin{equation}
v\mid _{\Gamma _{c}}=\chi g_{0},\text{ }\partial _{n}v\mid _{\Gamma
_{c}}=g_{0}\partial _{n}\chi +\chi g_{1},  \label{2.11_5}
\end{equation}%
\begin{equation}
v\left( x\right) =0,x\in \Omega _{c}\diagdown \Omega _{c+\varepsilon }.
\label{2.11_6}
\end{equation}%
Square both sides of (\ref{2.11_4}), multiply by $\varphi _{\lambda
}^{2}\left( x\right) $ and apply (\ref{2.6}). We obtain for all $\lambda
>\lambda _{0}$ and all $x\in \Omega _{c}$ 
\begin{eqnarray*}
&&C_{2}f^{2}\varphi _{\lambda }^{2}\left( x\right)
+C_{2}\sum\limits_{\left\vert \alpha \right\vert \leq 2}\left\vert D^{\alpha
}\left( 1-\chi \right) u\right\vert ^{2}\varphi _{\lambda }^{2}\left(
x\right) -\func{div}U \\
&\geq &C_{1}\lambda \left( 1-\frac{C_{1}}{\lambda }\right) \left( \nabla
v\right) ^{2}\varphi _{\lambda }^{2}\left( x\right) +C_{1}\lambda ^{3}\left(
1-\frac{C_{1}}{\lambda ^{3}}\right) v^{2}\varphi _{\lambda }^{2}\left(
x\right) .
\end{eqnarray*}%
Let $\lambda >\lambda _{1}:=\max \left( \lambda _{0},2C_{1}\right) .$ Then $%
C_{1}/\lambda <1/2.$Then with a different constant $C_{1}$%
\begin{eqnarray*}
&&C_{2}f^{2}\varphi _{\lambda }^{2}\left( x\right)
+C_{2}\sum\limits_{\left\vert \alpha \right\vert \leq 2}\left\vert D^{\alpha
}\left( 1-\chi \right) u\right\vert ^{2}\varphi _{\lambda }^{2}\left(
x\right) -\func{div}U \\
&\geq &C_{1}\lambda \left( \nabla u\right) ^{2}\varphi _{\lambda }^{2}\left(
x\right) +C_{1}\lambda ^{3}u^{2}\varphi _{\lambda }^{2}\left( x\right) , \\
\forall \lambda  &>&\lambda _{1},\forall x\in \Omega _{c}.
\end{eqnarray*}%
Integrate this inequality over $\Omega _{c}$ using Gauss' formula as well as
(\ref{2.1}), (\ref{2.3}), (\ref{2.8}), (\ref{2.11_3}), (\ref{2.11_5}) and (%
\ref{2.11_6}). We obtain%
\begin{eqnarray}
&&C_{2}e^{2\lambda m}\int\limits_{\Omega _{c}}f^{2}dx+C_{2}\lambda
^{3}e^{2\lambda m}\int\limits_{\Gamma _{c}}\left[ \left( \nabla g_{0}\right)
^{2}+g_{1}^{2}\right] dS_{x}+C_{2}\exp \left[ 2\lambda \left( c+2\varepsilon
\right) \right] \left\Vert u\right\Vert _{H^{2}\left( \Omega \right) }^{2} 
\notag \\
&\geq &\lambda \int\limits_{\Omega _{c}}\left( \nabla v\right) ^{2}\varphi
_{\lambda }^{2}dx+\lambda ^{3}\int\limits_{\Omega _{c}}v^{2}\varphi
_{\lambda }^{2}dx.  \label{2.12}
\end{eqnarray}%
Since $\Omega _{c+3\varepsilon }\subset \Omega _{c+2\varepsilon }\subset
\Omega _{c}$, then strengthening inequality (\ref{2.12}) and using (\ref%
{2.11_3}), we obtain%
\begin{eqnarray*}
&&C_{2}e^{2\lambda m}\int\limits_{\Omega _{c}}f^{2}dx+C_{2}\lambda
^{3}e^{2\lambda m}\int\limits_{\Gamma _{c}}\left[ \left( \nabla g_{0}\right)
^{2}+g_{1}^{2}\right] dS_{x}+C_{2}\exp \left[ 2\lambda \left( c+2\varepsilon
\right) \right] \left\Vert u\right\Vert _{H^{2}\left( \Omega \right) }^{2} \\
&\geq &\lambda \exp \left[ 2\lambda \left( c+3\varepsilon \right) \right]
\int\limits_{\Omega _{c+3\varepsilon }}\left[ \left( \nabla u\right)
^{2}+u^{2}\right] dx.
\end{eqnarray*}%
Using density arguments, we can relax now the $C^{2}-$smoothness of the
function $u$ and can claim that this inequality is also valid for $u\in
H^{2}\left( \Omega \right) .$ Dividing both sides of this inequality by $%
\lambda \exp \left[ 2\lambda \left( c+3\varepsilon \right) \right] $, we
conclude that there exists a number

$\lambda _{2}=\lambda _{2}\left( \varepsilon ,m,B,A_{0},\Omega _{c}\right)
>\lambda _{1}$ such that for all $\lambda >\lambda _{2}$ 
\begin{equation*}
\left\Vert u\right\Vert _{H^{1}\left( \Omega _{c+3\varepsilon }\right)
}^{2}\leq C_{2}\exp \left[ -2\lambda \varepsilon \right] \left\Vert
u\right\Vert _{H^{2}\left( \Omega _{c}\right) }^{2}+C_{2}\left( \left\Vert
g_{0}\right\Vert _{H^{1}\left( \Gamma _{c}\right) }^{2}+\left\Vert
g_{1}\right\Vert _{L_{2}\left( \Gamma _{c}\right) }^{2}+\left\Vert
f\right\Vert _{L_{2}\left( \Omega _{c}\right) }^{2}\right) e^{3\lambda m}.
\end{equation*}%
Hence, using (\ref{2.11_1}), we obtain 
\begin{equation}
\left\Vert u\right\Vert _{H^{1}\left( G_{c+2\varepsilon }\right) }^{2}\leq
C_{2}\left( \delta ^{2}e^{3\lambda m}+e^{-2\lambda \varepsilon }\left\Vert
u\right\Vert _{H^{2}\left( \Omega _{c}\right) }^{2}\right) .  \label{2.14}
\end{equation}%
We now balance two terms in the right hand side of (\ref{2.14}) via choosing 
$\lambda =\lambda \left( \delta \right) $ such that $\delta ^{2}e^{3\lambda
m}=e^{-2\lambda \varepsilon }.$ Hence, 
\begin{equation}
\lambda =\ln \left( \delta ^{-2\left( 3m+2\varepsilon \right) ^{-1}}\right) .
\label{2.15}
\end{equation}%
Hence, we should have $\delta \in \left( 0,\delta _{0}\right) ,$ where the
number $\delta _{0}=\delta _{0}\left( \varepsilon ,m,B,A_{0},\Omega
_{c}\right) $ is so small that

$\ln \left( \delta _{0}^{-2\left( 3m+2\varepsilon \right) ^{-1}}\right)
>\lambda _{2}.$ The target estimate (\ref{2.11_2}) follows from (\ref{2.14})
and (\ref{2.15}). $\square $

\textbf{Theorem 2.2} (uniqueness). \emph{Let conditions of Theorem 2.1 hold,
in (\ref{2.11}) }$g_{0}\left( x\right) \equiv g_{1}\left( x\right) \equiv
0,x\in \Gamma _{c}$\emph{\ and also }$f\left( x\right) \equiv 0.$\emph{\
Then }$u\left( x\right) \equiv 0$ \emph{\ for }$x\in \Omega _{c}.$

This theorem immediately follows from Theorem 2.1. To prove convergence of
minimizers of the Tikhonov functional to the correct solution (subsection
2.3), we need to replace the pointwise inequality (\ref{2.10}) with the
following integral inequality%
\begin{equation}
\left\Vert Au_{\delta }\right\Vert _{L_{2}\left( \Omega _{c}\right)
}^{2}\leq K\delta ^{2},K=const.\geq 1.  \label{2.150}
\end{equation}

\textbf{Theorem 2.3 }(H\"{o}lder stability estimate). \emph{Let the }$\delta
-$\emph{dependent family of functions} $u_{\delta }\in H^{2}\left( \Omega
_{c}\right) $\emph{\ satisfies inequality (\ref{2.150}) with the constant }$%
K $\emph{\ independent on} $\delta $\emph{.\ Assume that each function }$%
u_{\delta }$\emph{\ has zero boundary conditions (\ref{2.11})} \emph{and that%
} \emph{the Carleman estimate of Definition 2.1 is valid.\ Suppose that
there exists a sufficiently small number }$\varepsilon >0$\emph{\ such that }%
$\Omega _{c+3\varepsilon }\neq \varnothing $ \emph{and} $\Gamma
_{c+3\varepsilon }\neq \varnothing .$\ \emph{Then for the same numbers }$%
m,\beta $\emph{\ as} \emph{in Theorem 2.1} \emph{there exists a sufficiently
small number }$\delta _{0}=\delta _{0}\left( \varepsilon ,m,A,\Omega
_{c},K\right) \in \left( 0,1\right) $\emph{\ and a constant }$%
C_{3}=C_{3}\left( \varepsilon ,m,A,\Omega _{c},K\right) >0$\emph{\ such that
for all }$\delta \in \left( 0,\delta _{0}\right) $ \emph{the following H\"{o}%
lder stability estimate holds}%
\begin{equation*}
\left\Vert u_{\delta }\right\Vert _{H^{1}\left( \Omega _{c+3\varepsilon
}\right) }\leq C_{3}\left( 1+\left\Vert u_{\delta }\right\Vert _{H^{2}\left(
\Omega _{c}\right) }\right) \delta ^{\beta },\forall \delta \in \left(
0,\delta _{0}\right) .
\end{equation*}

\textbf{Proof}. In this proof $C_{3}=C_{3}\left( \varepsilon ,m,A,\Omega
_{c},K\right) >0$ denotes different positive constants depending on listed
parameters. Assume first that the function $u\in C^{2}\left( \overline{%
\Omega }_{c}\right) .$ We have 
\begin{equation*}
K\delta ^{2}e^{2\lambda m}\geq \int\limits_{\Omega _{c}}\left( Au\right)
^{2}\varphi _{\lambda }^{2}\left( x\right) dx\geq \int\limits_{\Omega
_{c}}\left( A_{0}u\right) ^{2}\varphi _{\lambda }^{2}\left( x\right)
dx-C_{3}\int\limits_{\Omega _{c}}\left( \left( \nabla u\right)
^{2}+u^{2}\right) \varphi _{\lambda }^{2}\left( x\right) dx.
\end{equation*}%
This is equivalent with%
\begin{equation*}
K\delta ^{2}e^{2\lambda m}+C_{3}\int\limits_{\Omega _{c}}\left( \left(
\nabla u\right) ^{2}+u^{2}\right) \varphi _{\lambda }^{2}\left( x\right)
dx\geq \int\limits_{\Omega _{c}}\left( A_{0}u\right) ^{2}\varphi _{\lambda
}^{2}\left( x\right) dx.
\end{equation*}%
The rest of the proof is similar with the proof of Theorem 2.1. The
replacement of $u\in C^{2}\left( \overline{\Omega }_{c}\right) $ with $u\in
H^{2}\left( \Omega _{c}\right) $ is done using density arguments. $\square $

\subsection{Regularization}

\label{sec:2.3}

\textbf{Cauchy Problem}. \emph{Find the function }$u\in H^{2}\left( \Omega
_{c}\right) $\emph{\ satisfying the following conditions} 
\begin{eqnarray}
Au &=&f\text{  in }\Omega _{c},  \label{2.102} \\
u &\mid &_{\Gamma _{c}}=g_{0}\left( x\right) ,\partial _{n}u\mid _{\Gamma
_{c}}=g_{1}\left( x\right) .  \label{2.103}
\end{eqnarray}

Assume that there exists a function $F\in H^{2}\left( \Omega _{c}\right) $
such that%
\begin{equation}
F\mid _{\Gamma _{c}}=g_{0}\left( x\right) ,\partial _{n}F\mid _{\Gamma
_{c}}=g_{1}\left( x\right) .  \label{2.1030}
\end{equation}%
We find an approximate solution of the problem (\ref{2.102}), (\ref{2.103})%
\emph{\ }as a minimizer of\emph{\ }the following Tikhonov functional with
the regularization parameter $\gamma \in \left( 0,1\right) ,$%
\begin{eqnarray}
J_{\gamma }\left( u\right) &=&\left\Vert Au-f\right\Vert _{L_{2}\left(
\Omega _{c}\right) }^{2}+\gamma \left\Vert u-F\right\Vert _{H^{2}\left(
\Omega _{c}\right) }^{2},u\in H^{2}\left( \Omega _{c}\right) ,  \label{2.104}
\\
&&\text{ subject to the Cauchy boundary data (\ref{2.103}).}  \label{2.105}
\end{eqnarray}%
In the regularization theory, such a minimizer is called \emph{regularized
solution, }see, e.g. \cite{BK,Tikh}. Thus, we regularize the problem (\ref%
{2.102}), (\ref{2.103}), which, at least in general, is ill-posed. First, we
prove the existence and uniqueness of the minimizer of the functional (\ref%
{2.104}) with conditions (\ref{2.105}).

\textbf{Theorem 2.4 }(existence).\textbf{\ }\emph{For every }$\gamma \in
\left( 0,1\right) $\emph{\ there exists unique minimizer }$u_{\gamma }\in
H^{2}\left( \Omega _{c}\right) $\emph{\ of the functional }$J_{\gamma
}\left( u\right) $\emph{\ and with a constant }$C_{4}=C_{4}\left( \Omega
_{c},A\right) >0$ \emph{the following estimate holds }%
\begin{equation}
\left\Vert u_{\gamma }\right\Vert _{H^{2}\left( \Omega _{c}\right) }\leq 
\frac{C_{4}}{\sqrt{\gamma }}\left( \left\Vert F\right\Vert _{H^{2}\left(
\Omega _{c}\right) }+\left\Vert f\right\Vert _{L_{2}\left( \Omega
_{c}\right) }\right) .  \label{2.107}
\end{equation}%
$.$\textbf{Proof}. In this proof $C_{4}=C_{4}\left( \Omega _{c},A\right) >0$
denotes different constant depending on listed parameters. Denote 
\begin{equation*}
H_{0,c}^{2}\left( \Omega _{c}\right) =\left\{ v\in H^{2}\left( \Omega
_{c}\right) :v\mid _{\Gamma _{c}}=\partial _{n}v\mid _{\Gamma
_{c}}=0\right\} .
\end{equation*}%
Let $v=u-F.$ Then $v\in H_{0,c}^{2}\left( \Omega _{c}\right) .$By (\ref%
{2.1030}), (\ref{2.104}) and (\ref{2.105}) we should minimize the following
functional $\overline{J}_{\gamma }\left( v\right) $%
\begin{equation}
\overline{J}_{\gamma }\left( v\right) =\left\Vert Av+\left( AF-f\right)
\right\Vert _{L_{2}\left( \Omega _{c}\right) }^{2}+\gamma \left\Vert
v\right\Vert _{H^{2}\left( \Omega _{c}\right) }^{2},v\in H_{0,c}^{2}\left(
\Omega _{c}\right) .  \label{2.108}
\end{equation}%
If $v_{\gamma }\in H_{0,c}^{2}\left( \Omega _{c}\right) $ is a minimizer of
the functional (\ref{2.108}), then $u_{\gamma }=v_{\gamma }+F$ is a
minimizer of the functional (\ref{2.104}) satisfying conditions (\ref{2.105}%
). And vice versa: if $u_{\gamma }$ is a minimizer of the functional (\ref%
{2.104}) satisfying conditions (\ref{2.105}), then $v_{\gamma }=u_{\gamma
}-F\in H_{0,c}^{2}\left( \Omega _{c}\right) $ is a minimizer of the
functional (\ref{2.108}).

By the variational principle any minimizer $v_{\gamma }$ of the functional (%
\ref{2.108}) should satisfy the following condition%
\begin{equation}
\left( Av_{\gamma },Ah\right) +\gamma \left[ v_{\gamma },h\right] =\left(
Ah,f-AF\right) ,\forall h\in H_{0,c}^{2}\left( \Omega _{c}\right) ,
\label{2.109}
\end{equation}%
where $\left( ,\right) $ and $\left[ ,\right] $ are scalar products in $%
L_{2}\left( \Omega _{c}\right) $ and $H^{2}\left( \Omega _{c}\right) $
respectively. Denote%
\begin{equation}
\left\{ v,h\right\} _{\gamma }=\left( Av,Ah\right) +\gamma \left[ v,h\right]
,\forall v,h\in H_{0,c}^{2}\left( \Omega _{c}\right) .  \label{2.110}
\end{equation}%
Hence, $\left\{ v,h\right\} $ defines a new scalar product in the Hilbert
space $H_{0}^{2}\left( \Omega _{c}\right) $ and the corresponding norm $%
\left\{ v\right\} $ satisfies%
\begin{equation}
\sqrt{\gamma }\left\Vert v\right\Vert _{H^{2}\left( \Omega _{c}\right) }\leq
\left\{ v\right\} _{\gamma }\leq C_{4}\left\Vert v\right\Vert _{H^{2}\left(
\Omega _{c}\right) }.  \label{2.111}
\end{equation}%
Thus, the scalar product (\ref{2.110}) generates the new norm $\left\{
v\right\} _{\gamma },$ which is equivalent with the norm $\left\Vert
v\right\Vert _{H^{2}\left( \Omega _{c}\right) }.$ Hence, (\ref{2.109}) can
be rewritten as%
\begin{equation}
\left\{ v_{\gamma },h\right\} _{\gamma }=\left( Ah,f-AF\right) ,\forall h\in
H_{0,c}^{2}\left( \Omega _{c}\right) .  \label{2.112}
\end{equation}%
It follows from (\ref{2.111}) that 
\begin{equation}
\left\vert \left( Ah,f-AF\right) \right\vert \leq C_{4}\left( \left\Vert
F\right\Vert _{H^{2}\left( \Omega _{c}\right) }+\left\Vert f\right\Vert
_{L_{2}\left( \Omega _{c}\right) }\right) \left\{ h\right\} _{\gamma }.
\label{2.113}
\end{equation}%
Hence, the right hand side of (\ref{2.112}) can be considered as a bounded
linear functional defined on the space $H_{0}^{2}\left( \Omega _{c}\right) .$
Hence, by Riesz theorem there exists an element $w_{\gamma }=w_{\gamma
}\left( f-AF\right) $ such that $\left( Ah,f-AF\right) =\left\{ w_{\gamma
},h\right\} _{\gamma },\forall h\in H_{0,c}^{2}\left( \Omega _{c}\right) .$
This and (\ref{2.112}) imply that $\left\{ v_{\gamma },h\right\} _{\gamma
}=\left\{ w_{\gamma },h\right\} _{\gamma },\forall h\in H_{0,c}^{2}\left(
\Omega _{c}\right) .$ Hence, the minimizer $v_{\gamma }$ exists and $%
v_{\gamma }=w_{\gamma }.$ Also, by Riesz theorem and (\ref{2.113}) $\left\{
v_{\gamma }\right\} _{\gamma }\leq C_{4}\left( \left\Vert F\right\Vert
_{H^{2}\left( \Omega _{c}\right) }+\left\Vert f\right\Vert _{L_{2}\left(
\Omega _{c}\right) }\right) .$ Hence, the minimizer$v_{\gamma }$ is unique
and the left inequality (\ref{2.111}) implies (\ref{2.107}). $\square $

In the proof of Theorem 2.4 we have used only the variational principle and
Riesz theorem. However, the Carleman estimate (\ref{2.6}) was not used. We
use this estimate in Theorem 2.5, which establishes the convergence rate of
minimizers $u_{\gamma }$ to the exact solution, under suitable conditions.
Note that convergence is established in a subdomain $\Omega _{c+3\varepsilon
}\subset \Omega _{c},$ which is a little bit less than the original domain $%
\Omega _{c}.$ Contrary to this, we show in section 6 that for the hyperbolic
case convergence takes place in the whole domain of interest, which is
actually the time cylinder in that case.

Following one of concepts of Tikhonov (see, e.g. section 1.4 of \cite{BK}),
we assume that there exists an exact solution $u^{\ast }$ of the problem (%
\ref{2.102}), (\ref{2.103}) with the exact data $f^{\ast }\in L_{2}\left(
\Omega _{c}\right) ,u^{\ast }\mid _{\Gamma _{c}}=g_{0}^{\ast }\in
H^{1}\left( \Gamma _{c}\right) ,\partial _{n}u^{\ast }\mid _{\Gamma
_{c}}=g_{1}^{\ast }\in L_{2}\left( \Gamma _{c}\right) .$ By Theorem 2.2 the
exact solution $u^{\ast }$ is unique. Because of the existence of $u^{\ast
}, $ there also exists an exact function $F^{\ast }\in H^{2}\left( \Omega
_{c}\right) $ satisfying boundary conditions (\ref{2.1030}), in which
functions $g_{0},g_{1}$ are replaced with functions $g_{0}^{\ast
},g_{1}^{\ast }.$ Here is an example of such a function $F^{\ast }.$ Let the
function $\rho \in C^{2}\left( \overline{\Omega }_{c}\right) $ be such that $%
\rho \left( x\right) =1$ in a small neighborhood $N_{\sigma }\left( \Gamma
_{c}\right) =\left\{ x\in \Omega _{c}:dist\left( x,\Gamma _{c}\right)
<\sigma \right\} $ \ and $\rho \left( x\right) =0$ for $x\in \Omega
_{c}\diagdown N_{2\sigma }\left( \Gamma _{c}\right) ,$ where $\sigma >0$ is
a sufficiently small number. Then $F^{\ast }$ can be constructed as $F^{\ast
}\left( x\right) =\rho \left( x\right) u^{\ast }\left( x\right) .$ Let $%
\delta >0$ be a sufficiently small number characterizing the error in the
data. We assume that 
\begin{equation}
\left\Vert f^{\ast }-f\right\Vert _{L_{2}\left( \Omega _{c}\right)
},\left\Vert g_{0}^{\ast }-g_{0}\right\Vert _{H^{1}\left( \Gamma _{c}\right)
},\left\Vert g_{1}^{\ast }-g_{1}\right\Vert _{L_{2}\left( \Gamma _{c}\right)
}\leq \delta ,\left\Vert F^{\ast }-F\right\Vert _{H^{2}\left( \Omega
_{c}\right) }\leq \delta .  \label{2.114}
\end{equation}

\textbf{Theorem 2.5} (convergence rate). \emph{Assume that the Carleman
estimate of Definition 2.1 holds, conditions (\ref{2.1030}) and (\ref{2.114}%
) are valid and let the regularization parameter }$\gamma =\gamma \left(
\delta \right) =\delta ^{2\alpha }$\emph{, where }$\alpha =const.\in \left(
0,1\right] .$\emph{\ Suppose that there exists a sufficiently small number }$%
\varepsilon >0$\emph{\ such that }$\Omega _{c+3\varepsilon }\neq \varnothing 
$ \emph{and} $\Gamma _{c+3\varepsilon }\neq \varnothing .$\emph{\ Let
numbers }$m,\beta $\emph{\ be the same as in Theorem 2.1. Then there exists
a sufficiently small number }$\delta _{0}=\delta _{0}\left( \varepsilon
,m,A,\Omega _{c}\right) \in \left( 0,1\right) $\emph{\ and a constant }$%
C_{5}=C_{5}\left( \varepsilon ,m,A,\Omega _{c}\right) >0$\emph{\ such that
if }$\delta \in \left( 0,\delta _{0}^{1/\alpha }\right) $,\emph{\ then the
following convergence rate is valid }%
\begin{equation}
\left\Vert u_{\gamma }-u^{\ast }\right\Vert _{H^{1}\left( \Omega
_{c+3\varepsilon }\right) }\leq C_{5}\left( 1+\left\Vert u^{\ast
}\right\Vert _{H^{2}\left( \Omega _{c}\right) }\right) \delta ^{\alpha \beta
},\forall \delta \in \left( 0,\delta _{0}\right) ,  \label{2.1140}
\end{equation}%
\emph{where }$u_{\gamma \left( \delta \right) }$\emph{\ is the minimizer of
the functional (\ref{2.104}), (\ref{2.105}) which is guaranteed by Theorem
2.4.}

\textbf{Proof}. In this proof $C_{5}=C_{5}\left( \varepsilon ,m,A,\Omega
_{c}\right) >0$ denotes different positive constants depending on listed
parameters. Let $v^{\ast }=u^{\ast }-F^{\ast }.$ Then $v^{\ast }\in
H_{0,c}^{2}\left( \Omega _{c}\right) $ and $Av^{\ast }=f^{\ast }-AF^{\ast }.$
Hence, 
\begin{equation}
\left( Av^{\ast },Ah\right) +\gamma \left[ v^{\ast },h\right] =\left(
Ah,f^{\ast }-AF^{\ast }\right) +\gamma \left[ v^{\ast },h\right] ,\forall
h\in H_{0}^{2}\left( \Omega _{c}\right) .  \label{2.115}
\end{equation}%
Subtract identity (\ref{2.109}) from identity (\ref{2.115}) and denote $%
\widetilde{v}_{\gamma }=v^{\ast }-v_{\gamma },\widetilde{f}=f^{\ast }-f,%
\widetilde{F}=F^{\ast }-F.$ Then%
\begin{equation*}
\left( A\widetilde{v}_{\gamma },Ah\right) +\gamma \left[ \widetilde{v}%
_{\gamma },h\right] =\left( Ah,\widetilde{f}-A\widetilde{F}\right) +\gamma %
\left[ v^{\ast },h\right] ,\forall h\in H_{0}^{2}\left( \Omega _{c}\right) .
\end{equation*}%
Setting here $h:=\widetilde{v}_{\gamma },$ we obtain 
\begin{equation}
\left\Vert A\widetilde{v}_{\gamma }\right\Vert _{L_{2}\left( \Omega
_{c}\right) }^{2}+\gamma \left\Vert \widetilde{v}_{\gamma }\right\Vert
_{H^{2}\left( \Omega _{c}\right) }^{2}=\left( A\widetilde{v}_{\gamma },%
\widetilde{f}-A\widetilde{F}\right) +\gamma \left[ v^{\ast },\widetilde{v}%
_{\gamma }\right] .  \label{2.116}
\end{equation}%
Applying the Cauchy-Schwarz inequality to (\ref{2.116}), we obtain%
\begin{equation}
\left\Vert A\widetilde{v}_{\gamma }\right\Vert _{L_{2}\left( \Omega
_{c}\right) }^{2}+\gamma \left\Vert \widetilde{v}_{\gamma }\right\Vert
_{H^{2}\left( \Omega _{c}\right) }^{2}  \label{2.1160}
\end{equation}%
\begin{equation*}
\leq \frac{1}{2}\left\Vert A\widetilde{v}_{\gamma }\right\Vert _{L_{2}\left(
\Omega _{c}\right) }^{2}+\frac{1}{2}\left\Vert \widetilde{f}-A\widetilde{F}%
\right\Vert _{L_{2}\left( \Omega _{c}\right) }^{2}+\frac{\gamma }{2}%
\left\Vert v^{\ast }\right\Vert _{H^{2}\left( \Omega _{c}\right) }^{2}+\frac{%
\gamma }{2}\left\Vert \widetilde{v}_{\gamma }\right\Vert _{H^{2}\left(
\Omega _{c}\right) }^{2}.
\end{equation*}%
Hence, by (\ref{2.114}) 
\begin{equation}
\left\Vert A\widetilde{v}_{\gamma }\right\Vert _{L_{2}\left( \Omega
_{c}\right) }^{2}+\gamma \left\Vert \widetilde{v}_{\gamma }\right\Vert
_{H^{2}\left( \Omega _{c}\right) }^{2}\leq C_{5}\delta ^{2}+\gamma
\left\Vert v^{\ast }\right\Vert _{H^{2}\left( \Omega _{c}\right) }^{2}.
\label{2.117}
\end{equation}%
Since $\gamma =\delta ^{2\alpha },$ where $\alpha \in \left( 0,1\right] ,$
then $\delta ^{2}\leq \gamma .$ Hence, (\ref{2.117}) implies that%
\begin{equation}
\left\Vert \widetilde{v}_{\gamma }\right\Vert _{H^{2}\left( \Omega
_{c}\right) }\leq C_{5}\left( 1+\left\Vert v^{\ast }\right\Vert
_{H^{2}\left( \Omega _{c}\right) }\right) ,  \label{2.120}
\end{equation}%
\begin{equation}
\left\Vert A\widetilde{v}_{\gamma }\right\Vert _{L_{2}\left( \Omega
_{c}\right) }^{2}\leq C_{5}\left( 1+\left\Vert v^{\ast }\right\Vert
_{H^{2}\left( \Omega _{c}\right) }^{2}\right) \delta ^{2\alpha }.
\label{2.121}
\end{equation}%
Let $w_{\gamma }=\widetilde{v}_{\gamma }\left( 1+\left\Vert v^{\ast
}\right\Vert _{H^{2}\left( \Omega _{c}\right) }\right) ^{-1}.$ Then (\ref%
{2.120}), (\ref{2.121}) and Theorem 2.3 imply that 

$\left\Vert w_{\gamma }\right\Vert _{H^{1}\left( \Omega _{c+3\varepsilon
}\right) }\leq C_{5}\delta ^{\alpha \beta },\forall \delta \in \left(
0,\delta _{0}\right) .$ Therefore,%
\begin{equation}
\left\Vert \widetilde{v}_{\gamma }\right\Vert _{H^{1}\left( \Omega
_{c+3\varepsilon }\right) }\leq C_{5}\left( 1+\left\Vert v^{\ast
}\right\Vert _{H^{2}\left( \Omega _{c}\right) }\right) \delta ^{\alpha \beta
},\text{ }\forall \delta \in \left( 0,\delta _{0}\right) .  \label{2.118}
\end{equation}%
Next, since $\widetilde{v}_{\gamma }=\left( u_{\gamma }-u^{\ast }\right)
+\left( F^{\ast }-F\right) $ and since by (\ref{2.114}) $\left\Vert F^{\ast
}-F\right\Vert _{H^{1}\left( \Omega _{c+3\varepsilon }\right) }\leq \delta ,$
then the triangle inequality implies that 
\begin{equation}
\left\Vert \widetilde{v}_{\gamma }\right\Vert _{H^{1}\left( \Omega
_{c+3\varepsilon }\right) }\geq \left\Vert u_{\gamma }-u^{\ast }\right\Vert
_{H^{1}\left( \Omega _{c+3\varepsilon }\right) }-\left\Vert F^{\ast
}-F\right\Vert _{H^{1}\left( \Omega _{c+3\varepsilon }\right) }\geq
\left\Vert u_{\gamma }-u^{\ast }\right\Vert _{H^{1}\left( \Omega
_{c+3\varepsilon }\right) }-\delta .  \label{2.119}
\end{equation}%
Since numbers $\beta ,\delta \in \left( 0,1\right) $ and since $\alpha \in
\left( 0,1\right] ,$ then $\delta ^{\alpha \beta }>\delta .$ Thus, using (%
\ref{2.118}) and (\ref{2.119}), we obtain (\ref{2.1140}). $\square $

\section{Cauchy Problem for the Elliptic Equation}

\label{sec:3}

We now rewrite the operator $A$ in (\ref{2.100}) as%
\begin{eqnarray}
Lu &=&\dsum\limits_{i,j=1}^{n}a_{i,j}\left( x\right)
u_{x_{i}x_{j}}+\dsum\limits_{j=1}^{n}b_{j}\left( x\right)
u_{x_{j}}+b_{0}\left( x\right) u,x\in \Omega ,  \label{3.1} \\
L_{0}u &=&\dsum\limits_{i,j=1}^{n}a_{i,j}\left( x\right) u_{x_{i}x_{j}},
\label{3.2}
\end{eqnarray}%
where $a_{i,j}\left( x\right) =a_{j,i}\left( x\right) ,\forall i,j$ and $%
L_{0}$ is the principal part of the operator $L.$ As in (\ref{2.101}), we
assume that%
\begin{equation}
a_{i,j}\in C^{1}\left( \overline{\Omega }\right) ;b_{j},b_{0}\in C\left( 
\overline{\Omega }\right) .  \label{3.3}
\end{equation}%
The ellipticity of the operator $L$ deals only with its principal part $%
L_{0} $ and it means that there exist two constants $\mu _{1},\mu _{2}>0,\mu
_{1}\leq \mu _{2}$ such that%
\begin{equation}
\mu _{1}\left\vert \eta \right\vert ^{2}\leq
\dsum\limits_{i,j=1}^{n}a_{i,j}\left( x\right) \eta _{i}\eta _{j}\leq \mu
_{2}\left\vert \eta \right\vert ^{2},\forall x\in \overline{\Omega },\forall
\eta =\left( \eta _{1},...\eta _{n}\right) \in \mathbb{R}^{n}.  \label{3.4}
\end{equation}%
Let $\Theta \subset \partial \Omega $ be the part of the boundary $\partial
\Omega $, where the Cauchy data are given. Assume that the equation of $%
\Theta $ is $\Theta =\left\{ x\in \mathbb{R}^{n}:x_{1}=g\left(
x_{2},...,x_{n}\right) ,\left( x_{2},...,x_{n}\right) \in \Theta ^{\prime
}\subset \mathbb{R}^{n-1}\right\} $ and that the function $g\in C^{2}\left( 
\overline{\Theta }^{\prime }\right) .$ Here $\Theta ^{\prime }\subset 
\mathbb{R}^{n-1}$ is a bounded domain. Changing variables as $x=\left(
x_{1},x_{2},...,x_{n}\right) \Leftrightarrow \left( x_{1}^{\prime
},x_{2},...,x_{n}\right) ,$ where $x_{1}^{\prime }=x_{1}-g\left(
x_{2},...,x_{n}\right) ,$ we obtain that in new variables

$\Theta =\left\{ x\in \mathbb{R}^{n}:x_{1}=0,\left( x_{2},...,x_{n}\right)
\in \Theta ^{\prime }\right\} .$ Here we kept the same notation for $x_{1}$
as before: for the simplicity of notations. This change of variables does
not affect the property of the ellipticity of the operator $L$. Let $X>0$ be
a certain number. Denote $\overline{x}=\left( x_{2},...,x_{n}\right) .$
Thus, without any loss of generality, we assume that 
\begin{equation}
\Omega \subset \left\{ x_{1}>0\right\} ,\text{ }\Theta =\left\{ x\in \mathbb{%
R}^{n}:x_{1}=0,\left\vert \overline{x}\right\vert <X\right\} \subset
\partial \Omega .  \label{3.5}
\end{equation}

Let the function $f\in L_{2}\left( \Omega \right) .$ Consider the elliptic
equation, 
\begin{equation}
Lu=f\text{ \ in }\Omega .  \label{3.6}
\end{equation}

\textbf{Cauchy Problem for the Elliptic Equation}. \emph{Let the part }$%
\Theta $\emph{\ of the boundary }$\partial \Omega $\emph{\ be given by (\ref%
{3.5}). Find such a function }$u\in H^{2}\left( \Omega \right) $\emph{\ that
satisfies equation (\ref{3.6}) and has the following Cauchy data }$%
g_{0},g_{1}$ \emph{at }$\Theta $%
\begin{equation}
u\mid _{\Theta }=g_{0}\left( \overline{x}\right) ,u_{x_{1}}\mid _{\Theta
}=g_{1}\left( \overline{x}\right) .  \label{3.7}
\end{equation}

These are incomplete Cauchy data, since they are given only at a part of the
boundary of the domain $\Omega $ rather than at the whole boundary. Let $%
\lambda >1$ and $\nu >1$ be two large parameters, which we define later.
Consider two arbitrary numbers $a,c=const.\in \left( 0,1\right) ,$ where $%
a<c $. To introduce the Carleman estimate, consider functions $\psi \left(
x\right) $, $\varphi _{\lambda }\left( x\right) $ defined as%
\begin{equation*}
\psi \left( x\right) =x_{1}+\frac{\left\vert \overline{x}\right\vert ^{2}}{%
X^{2}}+a,\varphi _{\lambda }\left( x\right) =\exp \left( \lambda \psi ^{-\nu
}\right) .
\end{equation*}%
Then the analogs of (\ref{2.0})\emph{\ }and\emph{\ }$\Gamma _{c}$ are 
\begin{eqnarray}
\Omega _{c} &=&\left\{ x:x_{1}>0,x_{1}+\frac{\left\vert \overline{x}%
\right\vert ^{2}}{X^{2}}+a<c\right\} ,\xi _{c}=\left\{ x:x_{1}>0,x_{1}+\frac{%
\left\vert \overline{x}\right\vert ^{2}}{X^{2}}+a=c\right\} ,  \label{3.8} \\
\Gamma _{c} &=&\left\{ x:x_{1}=0,\left\vert \overline{x}\right\vert <\sqrt{%
c-a}X\right\} ,  \label{3.9} \\
\partial \Omega _{c} &=&\xi _{c}\cup \Gamma _{c}.  \label{3.90}
\end{eqnarray}%
We assume that $\overline{\Omega }_{c}\subseteq \Omega .$ By (\ref{3.5}) and
(\ref{3.9}) $\Gamma _{c}\subset \Theta .$ For a sufficiently small number $%
\varepsilon >0$ define the subdomain $\Omega _{c+3\varepsilon }\subset
\Omega _{c}$ as 
\begin{equation}
\Omega _{c+3\varepsilon }=\left\{ x:x_{1}>0,x_{1}+\frac{\left\vert \overline{%
x}\right\vert ^{2}}{X^{2}}+a<c-3\varepsilon \right\} ,\varepsilon \in \left(
0,\left( c-a\right) /3\right) .  \label{3.91}
\end{equation}%
Lemma 3.1 follows immediately from Lemma 3 of \S 1 of chapter 4 of the book 
\cite{LRS}.

\textbf{Lemma 3.1 }(Carleman estimate). \emph{There exist a sufficiently
large number }

$\nu _{0}=\nu _{0}\left( a,c,\mu _{1},\mu _{2},\max_{i,j}\left\Vert
a_{i,j}\right\Vert _{C^{1}\left( \overline{\Omega }_{c}\right) },X\right) >1$%
\emph{\ and a sufficiently large absolute constant }$\lambda _{0}>1$\emph{\
such that for all }$\nu \geq \nu _{0},\lambda \geq \lambda _{0}$\emph{\ and
for all functions }$u\in C^{2}\left( \overline{\Omega }_{c}\right) $\emph{\
the following pointwise Carleman estimate is valid for all }$x\in \Omega
_{c} $\emph{\ with a constant }$C=C\left( n,\max_{i,j}\left\Vert
a_{i,j}\right\Vert _{C^{1}\left( \overline{\Omega }_{c}\right) }\right) $%
\begin{eqnarray*}
\left( L_{0}u\right) ^{2}\varphi _{\lambda }^{2} &\geq &C\lambda \left\vert
\nabla u\right\vert ^{2}\varphi _{\lambda }^{2}+C\lambda ^{3}u^{2}\varphi
_{\lambda }^{2}+\func{div}U, \\
\left\vert U\right\vert &\leq &C\lambda ^{3}\left[ \left( \nabla u\right)
^{2}+u^{2}\right] \varphi _{\lambda }^{2}.
\end{eqnarray*}

This Carleman estimate allows us to construct the Tikhonov functional for
solving the Cauchy problem (\ref{3.6}), (\ref{3.7}).\emph{\ }First, we
construct an example of the function $F\in H^{2}\left( \Omega _{c}\right) $:
as in (\ref{2.1030}). Assume that functions 
\begin{equation}
g_{0},g_{1}\in H^{2}\left( \Gamma _{c}\right) ,  \label{3.10}
\end{equation}%
where $\Gamma _{c}$ is defined in (\ref{3.9}). Even though the minimal
smoothness requirement should probably be $g_{0}\in H^{1}\left( \Gamma
_{c}\right) ,g_{1}\in L_{2}\left( \Gamma _{c}\right) ,$ we still assume a
little bit higher smoothness (\ref{3.10}) here only for the sake of our
specific example of the function $F$. Let $\sigma \in \left( 0,\left(
c-a\right) /2\right) $ be a sufficiently small number. Consider the function 
$\rho \left( x_{1}\right) $ such that 
\begin{equation}
\rho \in C^{2}\left[ 0,c-a\right] ,\rho \left( x_{1}\right) =\left\{ 
\begin{array}{c}
1,x_{1}\in \left( 0,\sigma \right) , \\ 
0,x_{1}\in \left[ 2\sigma ,c-a\right] .%
\end{array}%
\right.  \label{3.11}
\end{equation}%
We construct the function $F$ as 
\begin{equation}
F\left( x\right) =\rho \left( x_{1}\right) g_{0}\left( \overline{x}\right)
-\rho \left( x_{1}\right) x_{1}g_{1}\left( \overline{x}\right) .
\label{3.12}
\end{equation}%
By (\ref{3.10})-(\ref{3.12}) the function $F\in H^{2}\left( \Omega
_{c}\right) $ and also by (\ref{3.7})-(\ref{3.10}) $F\mid _{\Gamma
_{c}}=g_{0}\left( \overline{x}\right) ,\partial _{n}F\mid _{\Gamma
_{c}}=g_{1}\left( \overline{x}\right) ,$ where $\partial _{n}=-\partial
_{x_{1}}.$ We now assume the existence of the exact solution $u^{\ast }$ of
the problem (\ref{3.6}), (\ref{3.7}) with the exact Cauchy data $g_{0}^{\ast
},g_{1}^{\ast }\in H^{2}\left( \Gamma _{c}\right) $ and the exact right hand
side $f^{\ast }.$ Next, we construct the function $F^{\ast }\in H^{2}\left(
\Omega _{c}\right) $ as above via replacing in (\ref{3.12}) $g_{0},g_{1}$
with $g_{0}^{\ast },g_{1}^{\ast }.$

We construct the direct analog of the Tikhonov functional (\ref{2.104}) with
boundary conditions (\ref{2.105}) as%
\begin{eqnarray}
J_{\gamma }^{L}\left( u\right) &=&\left\Vert Lu-f\right\Vert _{L_{2}\left(
\Omega _{c}\right) }^{2}+\gamma \left\Vert u-F\right\Vert _{H^{2}\left(
\Omega _{c}\right) }^{2},u\in H^{2}\left( \Omega _{c}\right) ,  \label{3.13}
\\
&&\text{ subject to the Cauchy boundary data }g_{0},g_{1}\text{ on }\Gamma
_{c}\text{,}  \label{3.14}
\end{eqnarray}%
where the function $F$ is defined in (\ref{3.12}). Theorem 3.1 follows
immediately from Theorems 2.4.

\textbf{Theorem 3.1}. \emph{For every }$\gamma \in \left( 0,1\right) $\emph{%
\ there exists unique minimizer }$u_{\gamma }\in H^{2}\left( \Omega
_{c}\right) $\emph{\ of the functional }$J_{\gamma }^{L}\left( u\right) $%
\emph{\ and with a constant }$C_{6}=C_{6}\left( \Omega _{c},L,\rho
_{1}\right) >0$ \emph{the following estimate holds }%
\begin{equation}
\left\Vert u_{\gamma }\right\Vert _{H^{2}\left( \Omega _{c}\right) }\leq 
\frac{C_{6}}{\sqrt{\gamma }}\left( \left\Vert F\right\Vert _{H^{2}\left(
\Omega _{c}\right) }+\left\Vert f\right\Vert _{L_{2}\left( \Omega
_{c}\right) }\right) .  \label{3.15}
\end{equation}

The convergence Theorem 3.2 follows immediately from Theorems 2.5, 3.1,
Lemma 3.1 and (\ref{3.8})-(\ref{3.91}).

\textbf{Theorem 3.2 }(convergence rate). \emph{Assume that conditions (\ref%
{3.10})-(\ref{3.12}) are valid. Also, assume that conditions (\ref{2.114})
are satisfied. Let in (\ref{3.13}) the regularization parameter }$\gamma
=\gamma \left( \delta \right) =\delta ^{2\alpha }$\emph{, where }$\alpha
=const.\in \left( 0,1\right] .$\emph{\ Let }$\varepsilon >0$ \emph{be} \emph{%
a sufficiently small and the domain }$\Omega _{c+3\varepsilon }$\emph{\ be
as in (\ref{3.91}).\ Let the number }$m=a^{-\nu _{0}},$\emph{\ where }$\nu
_{0}=\nu _{0}\left( a,c,\mu _{1},\mu _{2},\max_{i,j}\left\Vert
a_{i,j}\right\Vert _{C^{1}\left( \overline{\Omega }_{c}\right) },X\right) >1$%
\emph{\ is the number of Lemma 3.1. Define the number }$\beta \in \left(
0,1\right) $\emph{\ as }$\beta =\left( 2\varepsilon \right) /\left(
3m+2\varepsilon \right) .$\emph{\ Then there exists a sufficiently small
number }$\delta _{0}=\delta _{0}\left( \varepsilon ,\nu _{0},L,\Omega
_{c},\rho _{1}\right) \in \left( 0,1\right) $\emph{\ and a constant }$%
C_{7}=C_{7}\left( \varepsilon ,\nu _{0},L,\Omega _{c},\rho _{1}\right) >0$%
\emph{\ such that if }$\delta \in \left( 0,\delta _{0}^{1/\alpha }\right) $%
\emph{, then the following convergence rate is valid }%
\begin{equation*}
\left\Vert u_{\gamma \left( \delta \right) }-u^{\ast }\right\Vert
_{H^{1}\left( \Omega _{c+3\varepsilon }\right) }\leq C_{7}\left(
1+\left\Vert u^{\ast }\right\Vert _{H^{2}\left( \Omega _{c}\right) }\right)
\delta ^{\alpha \beta },\forall \delta \in \left( 0,\delta _{0}\right) ,
\end{equation*}%
\emph{where }$u_{\gamma \left( \delta \right) }$\emph{\ is the minimizer of
the functional (\ref{3.13}), (\ref{3.14}), which is guaranteed by Theorem
3.1. }

\section{Parabolic Equation With the Lateral Cauchy Data}

\label{sec:4}

Let $G\subset \mathbb{R}^{n}$ be a bounded domain with a piecewise smooth
boundary and $T=const.>0.$ Denote $G_{T}=G\times \left( -T,T\right) .$ Let $%
L_{par}$ be the elliptic operator of the second order in $G_{T},$ which we
define the same way as the operator $L$ in (\ref{3.1})-(\ref{3.4}) with the
only difference that now its coefficients depend on both $x$ and $t$ and the
domain $\Omega $ is replaced with the domain $G_{T}.$ Let $L_{0,par}$ be the
similarly defined principal part of the operator $L_{par},$ see (\ref{3.2}).
For brevity we are not rewriting (\ref{3.1})-(\ref{3.4}). Next, we define
the parabolic operator as $P=\partial _{t}-L_{par}$ and the principal part
of $P$ is $P_{0}=\partial _{t}-L_{0,par}.$ Let $\Theta \subset \partial G$, $%
\Theta \in C^{2}$ be the subsurface of the boundary $\partial G$ with the
same properties as ones in section 3.\ Denote $\Theta _{T}=\Theta \times
\left( -T,T\right) .$ Without a loss of generality we assume that for a
certain number $X>0$ 
\begin{equation}
G\subset \left\{ x_{1}>0\right\} ,\Theta =\left\{ x\in \mathbb{R}%
^{n}:x_{1}=0,\left\vert \overline{x}\right\vert <X\right\} \subset \partial
G.  \label{4.1}
\end{equation}

Let the function $f\left( x,t\right) \in L_{2}\left( G_{T}\right) .$
Consider the parabolic equation 
\begin{equation}
Pu:=u_{t}-L_{par}u=f\text{ \ in }G_{T}.\text{ }  \label{4.2}
\end{equation}

\textbf{Cauchy Problem with the Lateral Data for the Parabolic Equation}. 
\emph{Let the part }$\Theta $\emph{\ of the boundary }$\partial G$\emph{\ be
given by equation (\ref{4.1}). Find such a function }$u\in H^{2}\left(
G_{T}\right) $\emph{\ that satisfies equation (\ref{4.2}) and has the
following lateral Cauchy data }$g_{0},g_{1}$ \emph{at }$\Theta _{T}$%
\begin{equation}
u\mid _{\Theta _{T}}=g_{0}\left( \overline{x},t\right) ,u_{x_{1}}\mid
_{\Theta _{T}}=g_{1}\left( \overline{x},t\right) .  \label{4.3}
\end{equation}

We are using here the smoothness $u\in H^{2}\left( G_{T}\right) $, which is
a little bit higher than possibly the minimal required smoothness $u\in
H^{2,1}\left( G_{T}\right) ,$ because we need this smoothness in Lemma 4.2.
We introduce the Carleman estimate similarly with section 3. Let $\lambda >1$
and $\nu >1$ be two large parameters, which we define later. Consider two
arbitrary numbers $a,c=const.\in \left( 0,1\right) ,$ where $a<c$. Consider
functions $\psi \left( x,t\right) $, $\varphi _{\lambda }\left( x,t\right) $
defined as%
\begin{equation*}
\psi \left( x,t\right) =x_{1}+\frac{\left\vert \overline{x}\right\vert ^{2}}{%
X^{2}}+\frac{t^{2}}{T^{2}}+a,\text{ }\varphi _{\lambda }\left( x,t\right)
=\exp \left( \lambda \psi ^{-\nu }\right) .
\end{equation*}%
Analogs of (\ref{3.8})-(\ref{3.91})\emph{\ }are 
\begin{eqnarray}
\Omega _{c} &=&\left\{ \left( x,t\right) :x_{1}>0,x_{1}+\frac{\left\vert 
\overline{x}\right\vert ^{2}}{X^{2}}+\frac{t^{2}}{T^{2}}+a<c\right\} ,
\label{4.4} \\
\xi _{c} &=&\left\{ \left( x,t\right) :x_{1}>0,x_{1}+\frac{\left\vert 
\overline{x}\right\vert ^{2}}{X^{2}}+\frac{t^{2}}{T^{2}}+a=c\right\} ,
\label{4.40} \\
\Gamma _{c} &=&\left\{ \left( x,t\right) :x_{1}=0,\frac{\left\vert \overline{%
x}\right\vert ^{2}}{X^{2}}+\frac{t^{2}}{T^{2}}<c-a\right\} ,  \label{4.5} \\
\partial \Omega _{c} &=&\xi _{c}\cup \Gamma _{c},  \label{4.6} \\
\Omega _{c+3\varepsilon } &=&\left\{ \left( x,t\right) :x_{1}>0,x_{1}+\frac{%
\left\vert \overline{x}\right\vert ^{2}}{X^{2}}+\frac{t^{2}}{T^{2}}%
+a<c-3\varepsilon \right\} ,\varepsilon \in \left( 0,\left( c-a\right)
/3\right) .  \label{4.7}
\end{eqnarray}%
We assume that $\Omega _{c}\subset \overline{G}_{T}.$ By (\ref{4.1}) and (%
\ref{4.6}) $\Gamma _{c}\subset \Theta _{T}.$ Lemma 4.1 follows immediately
from Lemma 3 of \S 1 of chapter 4 of the book \cite{LRS}.

\textbf{Lemma 4.1 }(Carleman estimate). \emph{There exist a sufficiently
large number }

$\nu _{0}=\nu _{0}\left( a,c,\mu _{1},\mu _{2},\max_{i,j}\left\Vert
a_{i,j}\right\Vert _{C^{1}\left( \overline{\Omega }_{c}\right) },X,T\right)
>1$\emph{\ and a sufficiently large absolute constant }$\lambda _{0}>1$\emph{%
\ such that for all }$\nu \geq \nu _{0},\lambda \geq \lambda _{0}$\emph{\
and for all functions }$u\in C^{2,1}\left( \overline{\Omega }_{c}\right) $%
\emph{\ the following pointwise Carleman estimate is valid for all }$\left(
x,t\right) \in \Omega _{c}$\emph{\ with a constant }$C=C\left(
n,\max_{i,j}\left\Vert a_{i,j}\right\Vert _{C^{1}\left( \overline{\Omega }%
_{c}\right) }\right) $%
\begin{eqnarray*}
\left( P_{0}u\right) ^{2}\varphi _{\lambda }^{2} &\geq &C\lambda \left\vert
\nabla u\right\vert ^{2}\varphi _{\lambda }^{2}+C\lambda ^{3}u^{2}\varphi
_{\lambda }^{2}+\func{div}U+V_{t}, \\
\left\vert U\right\vert ,\left\vert V\right\vert &\leq &C\lambda ^{3}\left[
\left( \nabla u\right) ^{2}+u_{t}^{2}+u^{2}\right] \varphi _{\lambda }^{2}.
\end{eqnarray*}

Similarly with (\ref{3.10}), let functions 
\begin{equation}
g_{0},g_{1}\in H^{2}\left( \Gamma _{c}\right)  \label{4.8}
\end{equation}%
with the same comments about a little bit higher smoothness requirements (%
\ref{4.8}) as in lines below (\ref{3.10}). In (\ref{4.8}) $\Gamma _{c}$ is
the same as in (\ref{4.5}). Let the number $\sigma $ be the same as in
section 3 and $\rho \left( x_{1}\right) $ be the function defined in (\ref%
{3.11}). We now construct the function $F\left( x,t\right) \in H^{2}\left(
\Omega _{c}\right) $ as 
\begin{equation}
F\left( x,t\right) =\rho \left( x_{1}\right) g_{0}\left( \overline{x}%
,t\right) -\rho \left( x_{1}\right) x_{1}g_{1}\left( \overline{x},t\right) .
\label{4.9}
\end{equation}%
By (\ref{4.5}), (\ref{4.8}) and (\ref{4.9}) $F\in H^{2}\left( \Omega
_{c}\right) .$ In addition, $F\mid _{\Gamma _{c}}=g_{0}\left( \overline{x}%
,t\right) ,\partial _{n}F\mid _{\Gamma _{c}}=g_{1}\left( \overline{x}%
,t\right) ,$ where $\partial _{n}=-\partial _{x_{1}}.$ Again, we assume the
existence of the exact solution $u^{\ast }$ of the problem (\ref{4.2}), (\ref%
{4.3}) with the exact Cauchy data $g_{0}^{\ast },g_{1}^{\ast }$ and with the
exact right hand side $f^{\ast }$ and we construct the function $F^{\ast
}\in H^{2,1}\left( \Omega _{c}\right) $ similarly.

Given the function $F$ in (\ref{4.9}), the Tikhonov functional is now
constructed as 
\begin{eqnarray}
J_{\gamma }\left( u\right) &=&\left\Vert Pu-f\right\Vert _{L_{2}\left(
\Omega _{c}\right) }^{2}+\gamma \left\Vert u-F\right\Vert _{H^{2}\left(
\Omega _{c}\right) }^{2},u\in H^{2}\left( \Omega _{c}\right) ,  \label{4.10}
\\
&&\text{ subject to the lateral Cauchy data (\ref{4.3}).}  \label{4.11}
\end{eqnarray}%
Similarly with Theorems 3.1, 3.2 and using the Carleman estimate of Lemma
4.1, we obtain Theorems 4.1, 4.2. We rely in these theorems on (\ref{4.4})-(%
\ref{4.11}).

\textbf{Theorem 4.1}. \emph{For every }$\gamma \in \left( 0,1\right) $\emph{%
\ there exists unique minimizer }$u_{\gamma }\in H^{2}\left( \Omega
_{c}\right) $\emph{\ of the functional }$J_{\gamma }^{P}\left( u\right) $%
\emph{\ in (\ref{4.10}), (\ref{4.11}) and with a constant }$%
C_{8}=C_{8}\left( \Omega _{c},P,\rho _{1}\right) >0$ \emph{the following
estimate holds }%
\begin{equation}
\left\Vert u_{\gamma }\right\Vert _{H^{2}\left( \Omega _{c}\right) }\leq 
\frac{C_{8}}{\sqrt{\gamma }}\left( \left\Vert F\right\Vert _{H^{2}\left(
\Omega _{c}\right) }+\left\Vert f\right\Vert _{L_{2}\left( \Omega
_{c}\right) }\right) .  \label{4.12}
\end{equation}

\textbf{Theorem 4.2 }(convergence rate). \emph{Assume that conditions (\ref%
{2.114}) are satisfied. Let in (\ref{4.10}) the regularization parameter }$%
\gamma =\gamma \left( \delta \right) =\delta ^{2\alpha }$\emph{, where }$%
\alpha =const.\in \left( 0,1\right] .$\emph{\ Let }$\varepsilon >0$ \emph{be}
\emph{a sufficiently small number and the domain }$\Omega _{c+3\varepsilon }$%
\emph{\ be as in (\ref{4.7}).\ Let the number }$m=a^{-\nu _{0}},$\emph{\
where }

$\nu _{0}=\nu _{0}\left( a,c,\mu _{1},\mu _{2},\max_{i,j}\left\Vert
a_{i,j}\right\Vert _{C^{1}\left( \overline{\Omega }_{c}\right) },X,T\right)
>1$\emph{\ is the number of Lemma 4.1. Define the number }$\beta \in \left(
0,1\right) $\emph{\ as }$\beta =\left( 2\varepsilon \right) /\left(
3m+2\varepsilon \right) .$\emph{\ Then there exists a sufficiently small
number }$\delta _{0}=\delta _{0}\left( \varepsilon ,\nu _{0},P,\Omega
_{c},\rho _{1}\right) \in \left( 0,1\right) $\emph{\ and a constant }$%
C_{9}=C_{9}\left( \varepsilon ,\nu _{0},P,\Omega _{c},\rho _{1}\right) >0$%
\emph{\ such that if }$\delta \in \left( 0,\delta _{0}^{1/\alpha }\right) $%
\emph{, then the following convergence rate is valid }%
\begin{equation*}
\left\Vert u_{\gamma \left( \delta \right) }-u^{\ast }\right\Vert
_{H^{1}\left( \Omega _{c+3\varepsilon }\right) }\leq C_{9}\left(
1+\left\Vert u^{\ast }\right\Vert _{H^{2}\left( \Omega _{c}\right) }\right)
\delta ^{\alpha \beta },\forall \delta \in \left( 0,\delta _{0}\right) ,
\end{equation*}%
\emph{where }$u_{\gamma \left( \delta \right) }$\emph{\ is the minimizer of
the functional (\ref{4.10}), (\ref{4.10}), which is guaranteed by Theorem
4.1. }

\section{Parabolic Equation With The Reversed Time}

\label{sec:5}

While we had the lateral Cauchy data in section 4, now we have the data at $%
\left\{ t=0\right\} .$ Furthermore, instead of the above H\"{o}lder
stability estimates in subdomains, we obtain below the logarithmic stability
estimate for the function $u\left( x,T\right) $, which means an estimate in
the whole domain.\ Thus, we need to reformulate the general scheme of
section 2 and provide new proofs of analogs of Theorems 2.1-2.5. It should
be pointed out that if we would estimate the solution for $t\in \left(
0,T-\varepsilon \right) $ for a small $\varepsilon >0,$ rather than at $%
\left\{ t=T\right\} ,$ then we would have the H\"{o}lder stability, see
Theorem 1 in \S 2 of Chapter 4 of the book Lavrentiev, Romanov and
Shishatskii \cite{LRS} and estimate (\ref{5.171}) below. However, we follow
here a modified version of the paper of the author \cite{Klib2006} and get
the logarithmic stability in the whole domain this way. The Carleman
estimate of Lemma 3 of \cite{Klib2006} is a modification of the Carleman
estimate of Lemma 3 of \S 2 of Chapter 4 of \cite{LRS}. The same is true for
Lemma 5.3 below. The key element of this modification, which is absent in 
\cite{LRS}, is the first term in the third line of (\ref{5.11}). Indeed,
after the integration over $t\in \left( 0,T\right) ,$ this term provides a
positive integral involving $u^{2}$ over $\left\{ t=T\right\} :$ because we
choose in Lemma 5.3 $k+T<a_{0},$ where the number $a_{0}>0$ is sufficiently
small. That positive integral, in turn allows us to obtain the logarithmic
stability estimate for the problem considered in this section. Still, since 
\cite{Klib2006} is concerned with estimates of initial conditions of
parabolic PDEs with lateral Cauchy data, rather then with the parabolic
equation with reversed time, we need to modify results of that paper here.%
\newline
We refer to books of Isakov \cite{Is} and Payne \cite{Payne} for the
so-called \textquotedblleft logarithmic convexity" method, which provides H%
\"{o}lder stability estimates for solutions of parabolic equations with the
reversed time for $u\left( x,T-\varepsilon \right) $ in the case when the
elliptic operator of that parabolic equation is self adjoint. Also, exercise
3.1.2 of \cite{Is} guarantees the logarithmic stability for $u\left(
x,T\right) $ under the assumption that the norm $\left\Vert u\left(
x,t\right) \right\Vert _{L_{2}\left( \Omega \right) }$ is uniformly bounder
for $t\in \left[ 0,T\right] .$ We do not use this assumption here.

\subsection{Problem statement}

\label{sec:5.1}

Again, let $G\subset \mathbb{R}^{n}$ be a bounded domain with a piecewise
smooth boundary and $T=const.>0.$ We now denote $Q_{T}=G\times \left(
0,T\right) ,S_{T}=\partial G\times \left( 0,T\right) .$ Similarly with
section 4, let $L_{par}$ be the elliptic operator of the second order in $%
Q_{T},$ whose coefficients depend on $x,t$ and satisfy conditions (\ref{3.3}%
), (\ref{3.4}) where $\Omega $ is replaced with $Q_{T}$ and the dependence
on $x$ is replaced with the dependence on $x,t$. Let $L_{0,par}$ be the
principal part of the operator $L$, like in (\ref{3.2}). Let the function $%
f\left( x,t\right) \in L_{2}\left( Q_{T}\right) .$ Consider the parabolic
equation with the reversed time in $Q_{T},$ supplied by an initial condition
and a Dirichlet boundary condition, 
\begin{eqnarray}
u_{t}+L_{par}u &=&f\text{ \ in }Q_{T},u\in H^{2}\left( Q_{T}\right) ,
\label{5.1} \\
\text{ }u\left( x,0\right)  &=&g\left( x\right) ,  \label{5.2}
\end{eqnarray}%
\begin{equation}
u\mid _{S_{T}}=p\left( x,t\right) .  \label{5.3}
\end{equation}%
Even though $u\in H^{2,1}\left( Q_{T}\right) $ in (\ref{5.1}) sounds more
natural than $u\in H^{2}\left( Q_{T}\right) $, we still need this extra
smoothness in (\ref{5.1}) for the derivation of the stability estimate of
Theorem 5.1 as well as for the convergence rate in Theorem 5.3. In fact, the
method presented in this section 5 enables one to replace the Dirichlet
boundary condition (\ref{5.3}) with the Neumann boundary condition. However,
we are not doing this here for brevity. In the elliptic operator $L_{par},$
let $C\left( \overline{Q}_{T}\right) -$norms of coefficients at lower order
terms be bounded by a positive constant $M,$ 
\begin{equation}
\left\Vert b_{j}\right\Vert _{C\left( \overline{Q}_{T}\right) }\leq
M,j=0,...,n.  \label{5.30}
\end{equation}

\textbf{The Parabolic Problem With The Reversed Time. }\emph{Given
conditions (\ref{5.1})-(\ref{5.3}), find the function }$u\left( x,T\right) .$

It is well known that this problem is ill-posed. Indeed, consider, for
example the following problem%
\begin{eqnarray*}
v_{t}+v_{xx} &=&0,\left( x,t\right) \in \left( 0,\pi \right) \times \left(
0,T\right) , \\
v\left( x,0\right) &=&r\left( x\right) , \\
v\left( 0,t\right) &=&v\left( \pi ,t\right) =0.
\end{eqnarray*}%
Let $r_{n}$ be Fourier coefficients of the function $r\left( x\right) $ with
respect to the functions $\left\{ \sin nx\right\} _{n=1}^{n}.$ Then%
\begin{equation}
v\left( x,t\right) =\dsum\limits_{n=1}^{\infty }r_{n}\sin nxe^{n^{2}t}.
\label{5.4}
\end{equation}%
It follows from (\ref{5.4}) that small perturbations of the function $%
r\left( x\right) $ can cause large perturbations of the function $v\left(
x,t\right) .$ Also, the solution of this problem exists only for a rather
narrow set of functions $r\left( x\right) :$ for those, for which the series
(\ref{5.4}) converges. Likewise, the larger $t$ is, the more unstable the
solution is. The latter is reflected in the fact that the Carleman estimate
of Lemma 5.3 below enables us to use only small values of $T$.

\subsection{The Carleman estimate}

\label{sec:5.2}

Let $k=const.>0$ be the number which we choose in Lemma 5.3. For $\lambda >1$
we now define the Carleman Weight Function $\varphi _{\lambda }\left(
t\right) $ as%
\begin{equation}
\varphi _{\lambda }\left( t\right) =\left( k+t\right) ^{-\lambda },t>0.
\label{5.5}
\end{equation}%
Thus, level hypersurfaces of this function are hyperplanes $\left\{
t=const.\right\} .$ Lemmata 5.1 and 5.2 are reformulations of Lemmata 1 and
2 respectively of \cite{Klib2006}.\ Hence, we do not prove them for brevity.

\textbf{Lemma 5.1}. \emph{There exists a sufficiently large number }

$\lambda _{0}=\lambda _{0}\left( \mu _{1},\mu _{2},\max_{i,j}\left\Vert
a_{i,j}\right\Vert _{C^{1}\left( \overline{\Omega }_{c}\right) }\right) >1$%
\emph{\ and a constant }$C=C\left( \mu _{1},\mu _{2},\max_{i,j}\left\Vert
a_{i,j}\right\Vert _{C^{1}\left( \overline{\Omega }_{c}\right) }\right) >0$%
\emph{\ such that for all }$\lambda \geq \lambda _{0}$\emph{\ and for all
functions }$u\in C^{2,1}\left( \overline{Q}_{T}\right) $\emph{\ the
following estimate holds with the function }$\varphi _{\lambda }\left(
t\right) $\emph{\ from (\ref{5.5}) and for all }$\left( x,t\right) \in Q_{T}$%
\emph{\ }%
\begin{eqnarray}
\left( -u_{t}-L_{0,par}u\right) u\varphi _{\lambda }^{2} &\geq &\mu
_{1}\left\vert \nabla u\right\vert ^{2}\varphi _{\lambda }^{2}-\lambda
u^{2}\varphi _{\lambda }^{2}+\func{div}U_{1}+\frac{\partial }{\partial t}%
\left( -\frac{u^{2}}{2}\varphi _{\lambda }^{2}\right) ,  \label{5.6} \\
\left\vert U_{1}\right\vert  &\leq &C\left\vert u\right\vert \left\vert
\nabla u\right\vert \varphi _{\lambda }^{2}.  \label{5.7}
\end{eqnarray}

\textbf{Lemma 5.2}. \emph{For numbers }$\lambda _{0},C$\emph{\ of Lemma 5.1,}
\emph{for all }$\lambda \geq \lambda _{0}$\emph{\ and for all functions }$%
u\in C^{2,1}\left( \overline{Q}_{T}\right) $\emph{\ the following estimate
holds with the function }$\varphi _{\lambda }\left( t\right) $\emph{\ from (%
\ref{5.5}) for all }$\left( x,t\right) \in Q_{T}$%
\begin{equation*}
\left( u_{t}+L_{0,par}u\right) ^{2}\varphi _{\lambda }^{2}\geq -C\left\vert
\nabla u\right\vert ^{2}\varphi _{\lambda }^{2}+\lambda \left( k+t\right)
^{-2}u^{2}\varphi _{\lambda }^{2}+
\end{equation*}%
\begin{equation}
\frac{\partial }{\partial t}\left( \lambda \left( k+t\right)
^{-1}u^{2}\varphi _{\lambda }^{2}-\varphi _{\lambda
}^{2}\dsum\limits_{i,j=1}^{n}a_{i,j}u_{x_{i}}u_{x_{j}}\right) +\func{div}%
U_{2},  \label{5.9}
\end{equation}%
\begin{equation*}
\left\vert U_{2}\right\vert \leq C\left\vert u_{t}\right\vert \left\vert
\nabla u\right\vert \varphi _{\lambda }^{2}.
\end{equation*}

To obtain the Carleman estimate out of these two lemmata, we should combine
them. This is done in Lemma 5.3. Although the dependence of numbers $\lambda
_{1},\theta ,C_{1}$ on the number $M$ defined in (\ref{5.30}) is not
necessary in Lemma 5.3, we still include this dependence in its formulation
in (\ref{5.90}), since we need it in the proof of Theorem 5.1.

\textbf{Lemma 5.3 }(Carleman estimate). \emph{There exists a sufficiently
small number }

$a_{0}=a_{0}\left( \mu _{1},\mu _{2},\max_{i,j}\left\Vert a_{i,j}\right\Vert
_{C^{1}\left( \overline{\Omega }_{c}\right) },M\right) \in \left( 0,1\right) 
$\emph{\ such that if }$k+T<a_{0},$\emph{\ then there exists a sufficiently
large number }$\lambda _{1}=\lambda _{1}\left( \mu _{1},\mu
_{2},\max_{i,j}\left\Vert a_{i,j}\right\Vert _{C^{1}\left( \overline{\Omega }%
_{c}\right) },M\right) $\emph{\ and constants }$C=C\left( \mu _{1},\mu
_{2},\max_{i,j}\left\Vert a_{i,j}\right\Vert _{C^{1}\left( \overline{\Omega }%
_{c}\right) }\right) >0,C_{1}=C_{1}\left( \mu _{1},\mu
_{2},\max_{i,j}\left\Vert a_{i,j}\right\Vert _{C^{1}\left( \overline{\Omega }%
_{c}\right) }\right) >0$\emph{\ and }$\theta =\theta \left( \mu _{1},\mu
_{2},\max_{i,j}\left\Vert a_{i,j}\right\Vert _{C^{1}\left( \overline{\Omega }%
_{c}\right) },M\right) $\emph{\ \ such that for all} $\lambda \geq \lambda
_{1}$ and \emph{for all functions }$u\in C^{2,1}\left( \overline{Q}%
_{T}\right) $\emph{\ the following estimate holds with the function }$%
\varphi _{\lambda }\left( t\right) $\emph{\ from (\ref{5.5}) for all }$%
\left( x,t\right) \in Q_{T}$%
\begin{equation*}
\left( u_{t}+L_{0,par}u\right) ^{2}\varphi _{\lambda }^{2}\geq \frac{2}{3}%
\theta \mu _{1}\left\vert \nabla u\right\vert ^{2}\varphi _{\lambda
}^{2}+C\lambda u^{2}\varphi _{\lambda }^{2}+\func{div}U
\end{equation*}%
\begin{equation}
+\frac{\partial }{\partial t}\left( C\lambda \left( k+t\right) ^{-1}\left(
1-C_{1}\left( k+t\right) \right) u^{2}\varphi _{\lambda }^{2}-C\varphi
_{\lambda }^{2}\dsum\limits_{i,j=1}^{n}a_{i,j}u_{x_{i}}u_{x_{j}}\right) ,
\label{5.11}
\end{equation}%
\begin{equation*}
\left\vert U\right\vert \leq C\left( \left\vert u_{t}\right\vert +\left\vert
u\right\vert \right) \left\vert \nabla u\right\vert \varphi _{\lambda }^{2}.
\end{equation*}

\textbf{Proof}. Choose a number $\theta =\theta \left( \mu _{1},\mu
_{2},\max_{i,j}\left\Vert a_{i,j}\right\Vert _{C^{1}\left( \overline{\Omega }%
_{c}\right) },M\right) >0$ such that $\theta \mu _{1}>2C$. Multiply (\ref%
{5.6}) by $\theta $ and sum up with (\ref{5.11}). We obtain%
\begin{eqnarray}
&&\left( u_{t}+L_{0,par}u\right) ^{2}\varphi _{\lambda }^{2}-\theta \left(
u_{t}+L_{0,par}u\right) u\varphi _{\lambda }^{2}  \notag \\
&\geq &\frac{\theta \mu _{1}}{2}\left\vert \nabla u\right\vert ^{2}\varphi
_{\lambda }^{2}+\lambda \left( k+t\right) ^{-2}\left( 1-\theta \left(
k+t\right) ^{2}\right) u^{2}\varphi _{\lambda }^{2}  \label{5.12} \\
&&+\frac{\partial }{\partial t}\left[ \lambda \left( k+t\right) ^{-1}\left(
1-\frac{\theta }{2}\left( k+t\right) \right) u^{2}\varphi _{\lambda
}^{2}-\varphi _{\lambda
}^{2}\dsum\limits_{i,j=1}^{n}a_{i,j}u_{x_{i}}u_{x_{j}}\right] +\func{div}U. 
\notag
\end{eqnarray}%
In (\ref{5.12}) $U=\theta U_{1}+U_{2}.$ Hence, using estimate (\ref{5.7}) as
well as estimate of the third line of (\ref{5.9}), we obtain the estimate of
the third line of (\ref{5.11}) for $\left\vert U\left( x,t\right)
\right\vert .$ Choose the number $a_{0}$ so small that 
\begin{equation}
\theta a_{0}^{2}<1/2.  \label{5.120}
\end{equation}
Since $k+T<a_{0},$ then (\ref{5.12}) implies that%
\begin{eqnarray}
&&\left( u_{t}+L_{0,par}u\right) ^{2}\varphi _{\lambda }^{2}-\theta \left(
u_{t}+L_{0,par}u\right) u\varphi _{\lambda }^{2}  \notag \\
&\geq &\frac{\theta \mu _{1}}{2}\left\vert \nabla u\right\vert ^{2}\varphi
_{\lambda }^{2}+\frac{\lambda }{2a_{0}^{2}}u^{2}\varphi _{\lambda }^{2}
\label{5.13} \\
&&+\frac{\partial }{\partial t}\left[ \lambda \left( k+t\right) ^{-1}\left(
1-C_{1}\left( k+t\right) \right) u^{2}\varphi _{\lambda }^{2}-\varphi
_{\lambda }^{2}\dsum\limits_{i,j=1}^{n}a_{i,j}u_{x_{i}}u_{x_{j}}\right] +%
\func{div}U.  \notag
\end{eqnarray}%
Next, by Cauchy-Schwarz inequality%
\begin{equation}
\left( u_{t}+L_{0,par}u\right) ^{2}\varphi _{\lambda }^{2}-\theta \left(
u_{t}+L_{0,par}u\right) u\varphi _{\lambda }^{2}\leq \frac{3}{2}\left(
u_{t}+L_{0,par}u\right) ^{2}\varphi _{\lambda }^{2}+\frac{\theta ^{2}}{2}%
u^{2}\varphi _{\lambda }^{2}.  \label{5.14}
\end{equation}%
Replacing the left hand side of (\ref{5.13}) with the right hand side of (%
\ref{5.14}), we obtain (\ref{5.11}). $\square $

\subsection{Stability estimates}

\label{sec:5.3}

For a sufficiently small parameter $\delta \in \left( 0,1\right) $ consider
the family of functions $u_{\delta }$ satisfying the following conditions%
\begin{eqnarray}
\dint\limits_{Q_{T}}\left( \partial _{t}u_{\delta }+L_{par}u_{\delta
}\right) ^{2}dxdt &\leq &N\delta ^{2},u_{\delta }\in H^{2}\left(
Q_{T}\right) ,  \label{5.15} \\
\left\Vert u_{\delta }\left( x,0\right) \right\Vert _{L_{2}\left( \Omega
\right) } &\leq &\sqrt{N}\delta .  \label{5.16} \\
u_{\delta } &\mid &_{S_{T}}=0,  \label{5.17}
\end{eqnarray}%
where the constant $N>0$ is independent on $\delta $. Conditions (\ref{5.15}%
)-(\ref{5.17}) are generalizations of conditions (\ref{5.1})-(\ref{5.3}) for
the case when $p\left( x,t\right) \equiv 0$ and $L_{2}-$norms of functions $%
f,g$ are sufficiently small.

\textbf{Theorem 5.1 }(stability estimates). \emph{Consider the family of
functions }$u_{\delta }$\emph{\ satisfying conditions (\ref{5.15})-(\ref%
{5.17}). Let the number }$T$\emph{\ is so small that }$2T<a_{0},$\emph{\
where the sufficiently small number }$a_{0}=a_{0}\left( \mu _{1},\mu
_{2},\max_{i,j}\left\Vert a_{i,j}\right\Vert _{C^{1}\left( \overline{\Omega }%
_{c}\right) },M\right) \in \left( 0,1\right) $\emph{\ was defined in Lemma
5.3. Then there exists a sufficiently small number }$\delta _{0}=\delta
_{0}\left( L_{par}\right) \in \left( 0,1\right) $\emph{\ and a number }$%
C_{10}=C_{10}\left( \mu _{1},\mu _{2},\max_{i,j}\left\Vert
a_{i,j}\right\Vert _{C^{1}\left( \overline{\Omega }_{c}\right)
},M,N,Q_{T}\right) >0$\emph{\ such that the following logarithmic stability
estimate holds for functions }$u_{\delta }\left( x,T\right) $%
\begin{equation}
\left\Vert u_{\delta }\left( x,T\right) \right\Vert _{L_{2}\left( \Omega
\right) }\leq \frac{C_{10}}{\sqrt{\ln \left( \delta ^{-1}\right) }}\left(
1+\left\Vert u_{\delta }\right\Vert _{H^{2}\left( Q_{T}\right) }\right) .
\label{5.170}
\end{equation}%
\emph{For every }$\varepsilon \in \left( 0,T/2\right) $\emph{\ define the
number }$\beta =\beta \left( \varepsilon \right) $ \emph{as }%
\begin{equation}
\beta =\beta \left( \varepsilon \right) =-\frac{\ln \left( 1-\varepsilon
/T\right) }{2\ln \left( 1-\varepsilon /2T\right) }\in \left( 0,\frac{1}{2}%
\right) .  \label{1}
\end{equation}%
\emph{Then the following H\"{o}lder stability estimate holds} 
\begin{equation}
\left\Vert \nabla u_{\delta }\right\Vert _{L_{2}\left( Q_{T-\varepsilon
}\right) }+\left\Vert u_{\delta }\right\Vert _{L_{2}\left( Q_{T-\varepsilon
}\right) }\leq C_{10}\left( 1+\left\Vert u_{\delta }\right\Vert
_{H^{2}\left( Q_{T}\right) }\right) \delta ^{\beta }.  \label{5.171}
\end{equation}

\textbf{Proof}. In this proof $C=C\left( \mu _{1},\mu
_{2},\max_{i,j}\left\Vert a_{i,j}\right\Vert _{C^{1}\left( \overline{\Omega }%
_{c}\right) }\right) $ denotes different positive constants depending on
listed parameters and the number $k=const.\in \left( 0,T\right] $. By (\ref%
{5.5}) $\max_{\left[ 0,T\right] }\varphi _{\lambda }^{2}\left( t\right)
=k^{-2\lambda }.$ Hence, by (\ref{5.15})%
\begin{eqnarray}
Nk^{-2\lambda }\delta ^{2} &\geq &\dint\limits_{Q_{T}}\left( \partial
_{t}u_{\delta }+L_{par}u_{\delta }\right) ^{2}\varphi _{\lambda
}^{2}dxdt\geq   \label{5.18} \\
&&\dint\limits_{Q_{T}}\left( \partial _{t}u_{\delta }+L_{0,par}u_{\delta
}\right) ^{2}\varphi _{\lambda }^{2}dxdt-CM^{2}\dint\limits_{Q_{T}}\left(
\left\vert \nabla u_{\delta }\right\vert ^{2}+u_{\delta }^{2}\right) \varphi
_{\lambda }^{2}dxdt.  \notag
\end{eqnarray}%
Integrate (\ref{5.11}) over $Q_{T}$ with the function $u:=u_{\delta }$ in
it. Even though in (\ref{5.11}) $u\in C^{2,1}\left( \overline{Q}_{T}\right) $
while $u_{\delta }\in H^{2}\left( Q_{T}\right) ,$ this can be handled by
density arguments. It follows from the Gauss formula, the third line of (\ref%
{5.11}) and (\ref{5.17}) that the boundary integral over $S_{T}$ is zero in
this case. Hence, we obtain%
\begin{equation*}
\dint\limits_{Q_{T}}\left( \partial _{t}u_{\delta }+L_{0,par}u_{\delta
}\right) ^{2}\varphi _{\lambda }^{2}dxdt\geq \frac{2}{3}\theta \mu
_{1}\dint\limits_{Q_{T}}\left\vert \nabla u_{\delta }\right\vert ^{2}\varphi
_{\lambda }^{2}dxdt+C_{10}\lambda \dint\limits_{Q_{T}}u_{\delta }^{2}\varphi
_{\lambda }^{2}dxdt
\end{equation*}%
\begin{eqnarray*}
&&+\left( k+T\right) ^{-2\lambda }\dint\limits_{\Omega }\left( C\lambda
\left( k+T\right) ^{-1}\left( 1-C_{1}\left( k+T\right) \right) u_{\delta
}^{2}-C\dsum\limits_{i,j=1}^{n}a_{i,j}\partial _{x_{i}}u_{\delta }\partial
_{x_{j}}u\right) \left( x,T\right) dx \\
&&+k^{-2\lambda }\dint\limits_{\Omega }\left( -C\lambda k\left(
1-C_{1}k\right) u_{\delta }^{2}+C\dsum\limits_{i,j=1}^{n}a_{i,j}\partial
_{x_{i}}u_{\delta }\partial _{x_{j}}u\right) \left( x,0\right) dx.
\end{eqnarray*}%
Hence, by (\ref{5.18})%
\begin{equation*}
Nk^{-2\lambda }\delta ^{2}\geq \frac{2}{3}\theta \mu
_{1}\dint\limits_{Q_{T}}\left\vert \nabla u_{\delta }\right\vert ^{2}\varphi
_{\lambda }^{2}dxdt+C_{10}\lambda \dint\limits_{Q_{T}}u_{\delta }^{2}\varphi
_{\lambda }^{2}dxdt
\end{equation*}%
\begin{equation}
-CM^{2}\dint\limits_{Q_{T}}\left( \left\vert \nabla u_{\delta }\right\vert
^{2}+u_{\delta }^{2}\right) \varphi _{\lambda }^{2}dxdt  \label{5.20}
\end{equation}%
\begin{eqnarray*}
&&+\left( k+T\right) ^{-2\lambda }\dint\limits_{\Omega }\left( C\lambda
\left( k+T\right) ^{-1}\left( 1-C_{1}\left( k+T\right) \right) u_{\delta
}^{2}-C\dsum\limits_{i,j=1}^{n}a_{i,j}\partial _{x_{i}}u_{\delta }\partial
_{x_{j}}u\right) \left( x,T\right) dx \\
&&+k^{-2\lambda }\dint\limits_{\Omega }\left( -C\lambda k\left(
1-C_{1}k\right) u_{\delta }^{2}+C\dsum\limits_{i,j=1}^{n}a_{i,j}\partial
_{x_{i}}u_{\delta }\partial _{x_{j}}u\right) \left( x,0\right) dx.
\end{eqnarray*}%
Choose $\lambda _{2}=\lambda _{2}\left( L_{par}\right) \geq \lambda _{1}$ so
large that $C_{10}\lambda >2CM^{2}.$ Next, choose 

$\theta =\theta \left( \mu _{1},\mu _{2},\max_{i,j}\left\Vert
a_{i,j}\right\Vert _{C^{1}\left( \overline{\Omega }_{c}\right) },M\right) $
so large that $\left( \theta \mu _{1}\right) /3>CM^{2}.$ Next, choose $%
a_{0}>0$ so small that $C_{1}a_{0}<1/4$ and also (\ref{5.120}) would be
satisfied. Recalling that $2T<a_{0}$, $k\in \left( 0,T\right) ,$ using (\ref%
{5.16}) and (\ref{5.20}) and taking into account (\ref{3.4}), we obtain 
\begin{equation}
\dint\limits_{Q_{T}}\left( \partial _{t}u_{\delta }+L_{0,par}u_{\delta
}\right) ^{2}\varphi _{\lambda }^{2}dxdt\geq \frac{1}{3}\theta \mu
_{1}\dint\limits_{Q_{T}}\left\vert \nabla u_{\delta }\right\vert ^{2}\varphi
_{\lambda }^{2}dxdt+C\lambda \dint\limits_{Q_{T}}u_{\delta }^{2}\varphi
_{\lambda }^{2}dxdt  \label{5.21}
\end{equation}%
\begin{equation*}
+C\lambda \left( k+T\right) ^{-2\lambda -1}\left\Vert u_{\delta }\left(
x,T\right) \right\Vert _{L_{2}\left( \Omega \right) }^{2}-C\left( k+T\right)
^{-2\lambda }\left\Vert \nabla u\left( x,T\right) \right\Vert _{L_{2}\left(
\Omega \right) }^{2}-CN\lambda k^{-2\lambda }\delta ^{2}.
\end{equation*}%
Hence, (\ref{5.18}) and (\ref{5.21}) lead to the following estimate 
\begin{eqnarray}
&&CN\lambda k^{-2\lambda }\delta ^{2}+C\left( k+T\right) ^{-2\lambda
}\left\Vert \nabla u\left( x,T\right) \right\Vert _{L_{2}\left( \Omega
\right) }^{2}  \label{5.22} \\
&\geq &\frac{1}{3}\theta \mu _{1}\dint\limits_{Q_{T}}\left\vert \nabla
u_{\delta }\right\vert ^{2}\varphi _{\lambda }^{2}dxdt+C\lambda
\dint\limits_{Q_{T}}u_{\delta }^{2}\varphi _{\lambda }^{2}dxdt+C\lambda
\left( k+T\right) ^{-2\lambda -1}\left\Vert u_{\delta }\left( x,T\right)
\right\Vert _{L_{2}\left( \Omega \right) }^{2}.  \notag
\end{eqnarray}

First, we obtain the logarithmic stability estimate (\ref{5.170}). By the
trace theorem, there exists a positive constant $D=D\left( Q_{T}\right) $
such that $\left\Vert \nabla w\left( x,T\right) \right\Vert _{L_{2}\left(
\Omega \right) }^{2}\leq D\left\Vert w\right\Vert _{H^{2}\left( Q_{T}\right)
}^{2}.$ Hence, dividing both sides of (\ref{5.22}) by $C\lambda \left(
k+T\right) ^{-2\lambda -1}$, recalling that $k+T\leq 2T<a_{0}<1$ and
ignoring first two terms in the second line of (\ref{5.22}), we obtain 
\begin{equation}
\left\Vert u_{\delta }\left( x,T\right) \right\Vert _{L_{2}\left( \Omega
\right) }^{2}\leq C_{10}\left( 1+\frac{T}{k}\right) ^{2\lambda }\delta ^{2}+%
\frac{1}{\lambda }C_{10}\left\Vert u_{\delta }\right\Vert _{H^{2}\left(
Q_{T}\right) }^{2}.  \label{5.23}
\end{equation}%
Set $k=T$ and choose $\lambda =\lambda \left( \delta \right) $ such that $%
2^{2\lambda }=1/\delta .$ Hence, 
\begin{equation}
\lambda \left( \delta \right) =\ln \left( \frac{1}{\delta }\right)
^{1/\left( 2\ln 2\right) }.  \label{5.24}
\end{equation}%
Naturally, we assume that $\delta $ is so small that $\lambda \left( \delta
\right) \geq \lambda _{2}.$ Hence, (\ref{5.23}) and (\ref{5.24}) imply that 
\begin{equation*}
\left\Vert u_{\delta }\left( x,T\right) \right\Vert _{L_{2}\left( \Omega
\right) }^{2}\leq \frac{C_{10}}{\ln \left( \delta ^{-1}\right) }\left(
1+\left\Vert u_{\delta }\right\Vert _{H^{2}\left( Q_{T}\right) }^{2}\right) .
\end{equation*}%
This, in turn implies (\ref{5.170}).

We now prove the H\"{o}lder stability estimate (\ref{5.171}). Recall that we
have now $k=T$. Since $\varphi _{\lambda }^{2}\left( t\right) \geq \left(
2T-\varepsilon \right) ^{-2\lambda }$ for $t\in \left( 0,T-\varepsilon
\right) ,$ then ignoring the last term in the second line of (\ref{5.22}),
we obtain from (\ref{5.22}) with a different constant $C$%
\begin{equation}
\left\Vert \nabla u_{\delta }\right\Vert _{L_{2}\left( Q_{T-\varepsilon
}\right) }^{2}+\left\Vert u_{\delta }\right\Vert _{L_{2}\left(
Q_{T-\varepsilon }\right) }^{2}\leq C_{10}\left( 2-\frac{\varepsilon }{T}%
\right) ^{2\lambda }\delta ^{2}+C_{10}\left( 1-\frac{\varepsilon }{2T}%
\right) ^{2\lambda }\left\Vert u_{\delta }\right\Vert _{H^{2}\left(
Q_{T}\right) }^{2}.  \label{5.25}
\end{equation}%
Since $\varepsilon \in \left( 0,T/2\right) ,$ then $2-\varepsilon /T>1.$
Hence, assuming that $\delta $ is sufficiently small, we can choose a
different $\lambda =\lambda \left( \delta \right) $ such that $\left(
2-\varepsilon /T\right) ^{2\lambda }=1/\delta .$ Hence, 
\begin{equation}
\lambda \left( \delta \right) =\frac{1}{2\ln \left( 2-\varepsilon /T\right) }%
\ln \left( \frac{1}{\delta }\right) .  \label{5.26}
\end{equation}%
For every $\varepsilon \in \left( 0,T/2\right) ,$ we have $\left(
1-\varepsilon /\left( 2T\right) \right) ^{2\lambda \left( \delta \right)
}=\delta ^{2\beta },$ where $\beta =\beta \left( \varepsilon \right) \in
\left( 0,1/2\right) $ is the number defined in (\ref{1}). Hence, (\ref{5.25}%
) and (\ref{5.26}) imply that 
\begin{equation*}
\left\Vert \nabla u_{\delta }\right\Vert _{L_{2}\left( Q_{T-\varepsilon
}\right) }^{2}+\left\Vert u_{\delta }\right\Vert _{L_{2}\left(
Q_{T-\varepsilon }\right) }^{2}\leq C_{10}\left( 1+\left\Vert u_{\delta
}\right\Vert _{H^{2}\left( Q_{T}\right) }^{2}\right) \delta ^{2\beta }.\text{
\ \ \ }\square
\end{equation*}

Theorem 5.2 follows immediately from Theorem 5.1.

\textbf{Theorem 5.2} (uniqueness). \emph{There exists at most one solution
of the problem (\ref{5.1})-(\ref{5.3}).}

\subsection{Regularization}

\label{sec:5.4}

We now construct the Tikhonov functional for the problem (\ref{5.1})-(\ref%
{5.3}). Suppose that there exists a function $F\in H^{2}\left( Q_{T}\right) $
such that 
\begin{equation}
F\left( x,0\right) =g\left( x\right) ,F\mid _{S_{T}}=p\left( x,t\right) .
\label{5.27}
\end{equation}%
The Tikhonov functional for the problem (\ref{5.1})-(\ref{5.3}) is%
\begin{eqnarray}
J_{\gamma ,L_{par}}\left( u\right) &=&\left\Vert u_{t}+L_{par}u\right\Vert
_{L_{2}\left( Q_{T}\right) }^{2}+\gamma \left\Vert u-F\right\Vert
_{H^{2}\left( Q_{T}\right) }^{2},u\in H^{2}\left( Q_{T}\right) ,
\label{5.28} \\
&&\text{subject to conditions (\ref{5.3}) and (\ref{5.3}). }  \label{5.29}
\end{eqnarray}

To establish the convergence rate of minimizers, we again assume the
existence of the exact solution $u^{\ast }\in H^{2}\left( Q_{T}\right) $ of
the problem (\ref{5.1})-(\ref{5.3}). This solution satisfies conditions (\ref%
{5.1})-(\ref{5.3}) with the exact data $g^{\ast },p^{\ast },f^{\ast }.$
Hence, there exists a function $F^{\ast }\in H^{2}\left( Q_{T}\right) $
satisfying conditions (\ref{5.27}) in which $g$ and $p$ are replaced with $%
g^{\ast }$ and $p^{\ast }$ respectively. We assume that 
\begin{equation}
\left\Vert g-g^{\ast }\right\Vert _{L_{2}\left( G\right) }\leq \delta
,\left\Vert f-f^{\ast }\right\Vert _{L_{2}\left( Q_{T}\right) },\left\Vert
F-F^{\ast }\right\Vert _{H^{2}\left( Q_{T}\right) }\leq \delta .
\label{5.31}
\end{equation}

Theorem 5.3 follows immediately from Theorem 2.4. Theorem 5.4 follows
immediately from Theorem 2.5, Remark 2.1 and Theorem 5.1.

\textbf{Theorem 5.3} (existence). \emph{Suppose that there exists a function 
}$F\in H^{2}\left( Q_{T}\right) $\emph{\ satisfying conditions (\ref{5.27}).
Then for each }$\gamma >0$\emph{\ there exists unique minimizer }$u_{\gamma
}\in H^{2}\left( Q_{T}\right) $\emph{\ of the functional (\ref{5.28}), (\ref%
{5.29}). Furthermore, with a constant }$C_{11}=C_{11}\left(
L_{par},Q_{T}\right) $\emph{\ the following estimate holds}%
\begin{equation}
\left\Vert u_{\gamma }\right\Vert _{H^{2}\left( Q_{T}\right) }\leq \frac{%
C_{11}}{\sqrt{\gamma }}\left( \left\Vert F\right\Vert _{H^{2,1}\left(
Q_{T}\right) }+\left\Vert f\right\Vert _{L_{2}\left( Q_{T}\right) }\right) .
\label{5.32}
\end{equation}

\textbf{Theorem 5.4} (convergence rate). \emph{Assume that conditions (\ref%
{5.31}) hold. Let in (\ref{5.28}) the regularization parameter }$\gamma
=\gamma \left( \delta \right) =\delta ^{2\alpha }$\emph{, where }$\alpha
=const.\in \left( 0,1\right] .$\emph{\ Let the number }$T$\emph{\ is so
small that }$T<a_{0}/2,$\emph{\ where the sufficiently small number }

$a_{0}=a_{0}\left( \mu _{1},\mu _{2},\max_{i,j}\left\Vert a_{i,j}\right\Vert
_{C^{1}\left( \overline{\Omega }_{c}\right) },M\right) \in \left( 0,1\right) 
$ \emph{was defined in Lemma 5.3}. \emph{Let the number }$\varepsilon \in
\left( 0,T/2\right) $ \emph{and let }$\beta =\beta \left( \varepsilon
\right) \in \left( 0,1/2\right) $\emph{\ be the number defined in (\ref{1}).}
\emph{Then there exists a sufficiently small number }$\delta _{0}=\delta
_{0}\left( L_{par},a_{0},Q_{T}\right) \in \left( 0,1\right) $\emph{\ and a
constant }$C_{12}=C_{12}\left( L_{par},a_{0},Q_{T}\right) >0$\emph{\ such
that for all }$\delta \in \left( 0,\delta _{0}^{1/\alpha }\right) $\emph{\
the following convergence rates are valid}%
\begin{equation*}
\left\Vert u_{\gamma \left( \delta \right) }\left( x,T\right) -u^{\ast
}\left( x,T\right) \right\Vert _{L_{2}\left( \Omega \right) }\leq \frac{%
C_{12}}{\sqrt{\ln \left( \delta ^{-1}\right) }}\left( 1+\left\Vert u^{\ast
}\right\Vert _{H^{2}\left( Q_{T}\right) }\right) ,
\end{equation*}%
\emph{\ }%
\begin{equation*}
\left\Vert \nabla u_{\gamma \left( \delta \right) }-\nabla u^{\ast
}\right\Vert _{L_{2}\left( Q_{T-\varepsilon }\right) }+\left\Vert u_{\gamma
\left( \delta \right) }-u^{\ast }\right\Vert _{L_{2}\left( Q_{T-\varepsilon
}\right) }\leq C_{12}\left( 1+\left\Vert u_{\delta }\right\Vert
_{H^{2}\left( Q_{T}\right) }\right) \delta ^{\alpha \beta },
\end{equation*}%
\emph{where }$u_{\gamma \left( \delta \right) }$\emph{\ is the minimizer of
the functional (\ref{5.28}), (\ref{5.29}), which is guaranteed by Theorem
5.3. }

\section{Hyperbolic Equation With Lateral Cauchy Data}

\label{sec:6}

Results of this section were originate in the work of Klibanov and Malinsky 
\cite{KM} and were developed further in works of Klibanov with coauthors 
\cite{ClK,Kaz,KR,KT,KKKN,Ksurvey}. As in subsection 5.1, let $G\subset 
\mathbb{R}^{n}$ be a bounded domain with a piecewise smooth boundary $%
\partial G$ and $T=const.>0.$ Denote 
\begin{equation}
Q_{T}=G\times \left( 0,T\right) ,S_{T}=\partial G\times \left( 0,T\right)
,Q_{T}^{\pm }=G\times \left( -T,T\right) ,S_{T}^{\pm }=\partial G\times
\left( -T,T\right) .  \label{6.0}
\end{equation}%
In this section we obtain both the Lipschitz type stability (Theorem 6.1)
and the Lipschitz type convergence rate (Theorem 6.3) in the whole time
cylinder $Q_{T}^{\pm }$ rather than weaker H\"{o}lder type estimates in
subdomains, as in previous sections. As it was mentioned in Introduction,
corresponding numerical studies of \cite{ClK,KR,KKKN} have demonstrated a
good performance.

The Carleman estimate of Lemma 6.1 for the hyperbolic operator $%
L_{0,hyp}=a\left( x\right) \partial _{t}^{2}-\Delta $ was proved in Theorem
1.10.2 of \cite{BK}. Other forms of Carleman estimates for the hyperbolic
case can be found in, e.g. Theorem 3.4.1 of \cite{Is}, Theorem 2.2.4 of \cite%
{KT}, Lemma 2 of \S 4 of chapter 4 of \cite{LRS} and in Lemma 3.1 \cite{Trig}%
.

\subsection{Problem statement}

\label{sec:6.1}

Let numbers $a_{l},a_{u}>0$ and $a_{l}<a_{u}.$ For $x\in G,$ let the
function $a\left( x\right) $ satisfy the following conditions in $G$%
\begin{equation}
a\left( x\right) \in \left[ a_{l},a_{u}\right] ,a\in C^{1}\left( \overline{G}%
\right) .  \label{6.1}
\end{equation}%
In addition, we assume that there exists a point $x_{0}\in G$ such that%
\begin{equation}
\left( \nabla a\left( x\right) ,x-x_{0}\right) \geq \alpha =const.>0,\forall
x\in \overline{G},  \label{6.2}
\end{equation}%
where $\left( \cdot ,\cdot \right) $ denotes the scalar product in $\mathbb{R%
}^{n}$.\emph{\ }We need inequality (\ref{6.2}) for the validity of the
Carleman estimate of Lemma 6.1. Also, let functions%
\begin{equation}
b_{j}\left( x,t\right) \in C\left( \overline{Q}_{T}^{\pm }\right) ,j=0,...,n;%
\text{ }M=\max_{j}\left\Vert b_{j}\right\Vert _{C\left( \overline{Q}%
_{T}^{\pm }\right) }.  \label{6.21}
\end{equation}%
Let the function $f\in L_{2}\left( Q_{T}^{\pm }\right) $ and the function $u$
satisfies the following conditions%
\begin{eqnarray}
L_{hyp}u &=&a\left( x\right) u_{tt}-\Delta
u-\dsum\limits_{j=1}^{n}b_{j}\left( x,t\right) u_{x_{j}}-b_{0}\left(
x,t\right) u=f\left( x,t\right) \text{ in }Q_{T}^{\pm },u\in H^{2}\left(
Q_{T}^{\pm }\right) ,  \label{6.3} \\
u &\mid &_{S_{T}^{\pm }}=g_{0}\left( x,t\right) ,\partial _{n}u\mid
_{S_{T}^{\pm }}=g_{1}\left( x,t\right) .  \label{6.4}
\end{eqnarray}

\textbf{Cauchy Problem with the Lateral Data for the Hyperbolic Equation (%
\ref{6.3})}. Let coefficients of the hyperbolic operator $L_{hyp}$ satisfy
conditions (\ref{6.1})-(\ref{6.21}). Find the function $u$ satisfying
conditions (\ref{6.3}), (\ref{6.4}).

Let the number $\eta \in \left( 0,1\right) .$ Let $\lambda >1$ be a large
parameter$.$ Define functions 

$\xi \left( x,t\right) ,\varphi _{\lambda }\left( x,t\right) $ as%
\begin{equation}
\xi \left( x,t\right) =\left\vert x-x_{0}\right\vert ^{2}-\eta t^{2},\varphi
_{\lambda }\left( x,t\right) =\exp \left[ \lambda \xi \left( x,t\right) %
\right] .  \label{2.47}
\end{equation}%
Following (\ref{2.0}), for a number $c>0$ define the hypersurface $\xi _{c}$
and the domain $\Omega _{c}$ as%
\begin{equation}
\xi _{c}=\left\{ \left( x,t\right) \in Q_{T}^{\pm }:\xi \left( x,t\right)
=c,\right\} ,\text{ }\Omega _{c}=\left\{ \left( x,t\right) \in Q_{T}^{\pm
}:\xi \left( x,t\right) >c\right\} .  \label{2.471}
\end{equation}

\textbf{Lemma 6.1 }(Carleman estimate). \emph{Let }$n\geq 2$\emph{\ and
conditions (\ref{6.1}) be satisfied. Also, assume that there exists a point }%
$x_{0}\in G$\emph{\ such that (\ref{6.2}) holds. Let }$M$\emph{\ be the
number in (\ref{6.21}). Denote }$P=P\left( x_{0},G\right) =\max_{x\in 
\overline{G}}\left\vert x-x_{0}\right\vert .$\emph{\ Then there exists a
number }$\eta _{0}=\eta _{0}\left( G,P,a_{l},a_{u},\left\Vert \nabla
a\right\Vert _{C\left( \overline{G}\right) }\right) \in \left( 0,1\right) $ 
\emph{such that for any }$\eta \in \left( 0,\eta _{0}\right) $\emph{\ one
can choose a sufficiently large number }$\lambda _{0}=\lambda _{0}\left(
G,P,a_{l},a_{u},\left\Vert \nabla a\right\Vert _{C\left( \overline{G}\right)
},M,\eta _{0},c\right) >1$\emph{\ and the number }$C_{13}=C_{13}\left(
G,P,a_{l},a_{u},\left\Vert \nabla a\right\Vert _{C\left( \overline{G}\right)
},M,\eta _{0},c\right) >0$\emph{, such that for all }$u\in C^{2}\left( 
\overline{\Omega }_{c}\right) $\emph{\ and for all }$\lambda \geq \lambda
_{0}$\emph{\ the following pointwise Carleman estimate holds} 
\begin{equation}
\left( L_{hyp}u\right) ^{2}\varphi ^{2}\geq C_{13}\lambda \left( \left\vert
\nabla u\right\vert ^{2}+u_{t}^{2}\right) \varphi _{\lambda
}^{2}+C_{13}\lambda ^{3}u^{2}\varphi _{\lambda }^{2}+\func{div}U+V_{t}\text{%
, \emph{in} }\Omega _{c},  \label{2.49}
\end{equation}%
\emph{\ } 
\begin{eqnarray}
\left\vert U\right\vert &\leq &C_{13}\lambda ^{3}\left( \left\vert \nabla
u\right\vert ^{2}+u_{t}^{2}+u^{2}\right) \varphi _{\lambda }^{2},
\label{2.50} \\
\left\vert V\right\vert &\leq &C_{13}\lambda ^{3}\left[ \left\vert
t\right\vert \left( u_{t}^{2}+\left\vert \nabla u\right\vert
^{2}+u^{2}\right) +\left( \left\vert \nabla u\right\vert +\left\vert
u\right\vert \right) \left\vert u_{t}\right\vert \right] \varphi _{\lambda
}^{2}.  \label{2.51}
\end{eqnarray}%
\emph{\ } \emph{In particular, (\ref{2.51}) implies that if either }$u\left(
x,0\right) =0$\emph{\ or }$u_{t}\left( x,0\right) =0,$ \emph{then } 
\begin{equation}
V\left( x,0\right) =0.  \label{2.52}
\end{equation}%
\emph{In the case }$a\left( x\right) \equiv 1$\emph{\ one can choose any }$%
\eta \in \left( 0,1\right) $ \emph{and condition (\ref{6.2}) is not required.%
}

\subsection{Lipschitz stability estimate}

\label{sec:6.2}

We now obtain the Lipschitz stability estimate for a problem, which is more
general than the problem (\ref{6.3}), (\ref{6.4}). Let the function $u\in
H^{2}\left( Q_{T}\right) $ satisfies conditions (\ref{6.4}) and the
following inequality%
\begin{equation}
\dint\limits_{Q_{T}^{\pm }}\left( L_{hyp}u\right) ^{2}dxdt\leq K^{2}.
\label{6.5}
\end{equation}

\textbf{Theorem 6.1}. \emph{Let conditions (\ref{6.2})-(\ref{6.21}) be
satisfied}$.$ \emph{Suppose that }$T>P/\sqrt{\eta _{0}},$\emph{\ where }$%
\eta _{0}=\eta _{0}\left( G,P,a_{l},a_{u},\left\Vert \nabla a\right\Vert
_{C\left( \overline{G}\right) }\right) \in \left( 0,1\right) $\emph{\ is the
number of Lemma 6.1. Then there exists a constant }$C_{14}=C_{14}\left(
P,a_{l},a_{u},\left\Vert \nabla a\right\Vert _{C\left( \overline{G}\right)
},M,\eta _{0},Q_{T}^{\pm }\right) >0$\emph{\ such that for any function }$%
u\in H^{2}\left( Q_{T}^{\pm }\right) $\emph{\ satisfying conditions (\ref%
{6.4}), (\ref{6.5}) the following Lipschitz stability estimate holds}%
\begin{equation}
\left\Vert u\right\Vert _{H^{1}\left( Q_{T}^{\pm }\right) }\leq C_{14}\left(
\left\Vert g_{0}\right\Vert _{H^{1}\left( S_{T}^{\pm }\right) }+\left\Vert
g_{1}\right\Vert _{L_{2}\left( S_{T}^{\pm }\right) }+K\right) .  \label{6.6}
\end{equation}%
\emph{In the case }$a\left( x\right) \equiv 1$\emph{\ and one can choose }$%
\eta _{0}=1,$ \emph{and if in this case }$G=\left\{ \left\vert x\right\vert
<R\right\} ,$\emph{\ then one can choose }$T>R.$

\textbf{Proof}. In this proof $C_{14}=C_{14}\left( P,a_{l},a_{u},\left\Vert
\nabla a\right\Vert _{C\left( \overline{G}\right) },M,\eta _{0},Q_{T}^{\pm
}\right) >0$ denotes different positive constants depending on listed
parameters. Choose the number $\eta \in \left( 0,\eta _{0}\right) $ so close
to $\eta _{0}$ that $T>P/\sqrt{\eta }.$ Then (\ref{2.471}) implies that 
\begin{equation}
\Omega _{c}\subset Q_{T}^{\pm },\text{ }\overline{\Omega }_{c}\cap \left\{
t=\pm T\right\} =\varnothing .  \label{6.70}
\end{equation}

Let $m=\max_{\overline{\Omega }_{c}}\xi \left( x,t\right) .$ Then $%
m=\max_{x\in \overline{\Omega }_{c}}\left\vert x-x_{0}\right\vert ^{2}.$ We
have%
\begin{equation*}
\dint\limits_{Q_{T}^{\pm }}\left( L_{hyp}u\right)
^{2}dxdt=\dint\limits_{Q_{T}^{\pm }}\left( L_{hyp}u\right) ^{2}\varphi
_{\lambda }^{2}\varphi _{\lambda }^{-2}dxdt\geq e^{-2\lambda
m}\dint\limits_{Q_{T}^{\pm }}\left( L_{hyp}u\right) ^{2}\varphi _{\lambda
}^{2}dxdt.
\end{equation*}%
Hence, using (\ref{6.5}), we obtain%
\begin{equation}
\dint\limits_{Q_{T}^{\pm }}\left( L_{hyp}u\right) ^{2}\varphi _{\lambda
}^{2}dxdt\leq K^{2}e^{2\lambda m}.  \label{6.9}
\end{equation}%
By (\ref{6.210}) and (\ref{6.9})%
\begin{equation*}
K^{2}e^{2\lambda m}\geq \dint\limits_{Q_{T}^{\pm }}\left( L_{hyp}u\right)
^{2}\varphi _{\lambda }^{2}dxdt\geq \dint\limits_{Q_{T}^{\pm }}\left(
L_{0,hyp}u\right) ^{2}\varphi _{\lambda
}^{2}dxdt-C_{14}\dint\limits_{Q_{T}^{\pm }}\left( \left\vert \nabla
u\right\vert ^{2}+u^{2}\right) ^{2}\varphi _{\lambda }^{2}dxdt.
\end{equation*}%
Hence,%
\begin{equation}
\dint\limits_{Q_{T}^{\pm }}\left( L_{0,hyp}u\right) ^{2}\varphi _{\lambda
}^{2}dxdt\leq C_{14}\dint\limits_{Q_{T}^{\pm }}\left( \left\vert \nabla
u\right\vert ^{2}+u^{2}\right) ^{2}\varphi _{\lambda
}^{2}dxdt+K^{2}e^{2\lambda m}.  \label{6.10}
\end{equation}%
Let $\omega \left( x,t\right) $ be a function such that%
\begin{equation}
\omega \in C^{2}\left( \overline{Q}_{T}^{\pm }\right) ,\omega \left(
x,t\right) =\left\{ 
\begin{array}{c}
1,\left( x,t\right) \in \Omega _{c+2\varepsilon }, \\ 
0,\left( x,t\right) \in Q_{T}^{\pm }\diagdown \Omega _{c+\varepsilon }, \\ 
\in \left[ 0,1\right] ,\left( x,t\right) \in \Omega _{c+\varepsilon
}\diagdown \Omega _{c+2\varepsilon }.%
\end{array}%
\right.   \label{6.11}
\end{equation}%
Consider the function $v=\omega u.$ Then 
\begin{equation*}
L_{0,hyp}v=\omega L_{0,hyp}u+2\left( a\left( x\right) \omega
_{t}u_{t}-\nabla \omega \nabla u\right) +uL_{0,hyp}\omega .
\end{equation*}%
Hence, using (\ref{6.10}), we obtain%
\begin{equation}
\dint\limits_{Q_{T}^{\pm }}\left( L_{0,hyp}v\right) ^{2}\varphi _{\lambda
}^{2}dxdt\leq C_{14}\dint\limits_{Q_{T}^{\pm }}\left( \left\vert \nabla
u\right\vert ^{2}+u^{2}\right) ^{2}\varphi _{\lambda
}^{2}dxdt+K^{2}e^{2\lambda m}.  \label{6.12}
\end{equation}%
On the other hand, integrate (\ref{2.49}) over $\Omega _{c}$ for the
function $v,$ using (\ref{2.50}), (\ref{2.51}), (\ref{6.70}), Gauss formula
and the fact that by (\ref{2.471}) and (\ref{6.11}) $v\mid _{\xi _{c}}=0.$
We obtain 
\begin{eqnarray}
\dint\limits_{Q_{T}^{\pm }}\left( L_{0,hyp}v\right) ^{2}\varphi _{\lambda
}^{2}dxdt &\geq &\dint\limits_{\Omega _{c}}\left( L_{0,hyp}v\right)
^{2}\varphi _{\lambda }^{2}dxdt\geq C_{14}\dint\limits_{\Omega _{c}}\left(
\lambda \left( \left\vert \nabla v\right\vert ^{2}+v_{t}^{2}\right) +\lambda
^{3}v^{2}\right) \varphi _{\lambda }^{2}dxdt  \notag \\
&&-C_{14}\lambda ^{3}\dint\limits_{\partial \Omega _{c}\cap S_{T}^{\pm
}}\left( \left\vert \nabla v\right\vert ^{2}+v_{t}^{2}+v^{2}\right) \varphi
_{\lambda }^{2}dS.  \label{6.13}
\end{eqnarray}%
Next, by (\ref{6.4})%
\begin{equation}
-C_{14}\lambda ^{3}\dint\limits_{\partial \Omega _{c}\cap S_{T}^{\pm
}}\left( \left\vert \nabla v\right\vert ^{2}+v_{t}^{2}+v^{2}\right) \varphi
_{\lambda }^{2}dS\geq -C_{14}\lambda ^{3}e^{2\lambda m}\left( \left\Vert
g_{0}\right\Vert _{H^{1}\left( Q_{T}^{\pm }\right) }^{2}+\left\Vert
g_{1}\right\Vert _{L_{2}\left( Q_{T}^{\pm }\right) }^{2}\right) .
\label{6.14}
\end{equation}%
Also, since $\Omega _{c+2\varepsilon }\subset \Omega _{c}$ and by (\ref{6.11}%
), $v=u$ in $\Omega _{c+2\varepsilon }$ and then%
\begin{equation*}
C_{14}\dint\limits_{\Omega _{c}}\left( \lambda \left( \left\vert \nabla
v\right\vert ^{2}+v_{t}^{2}\right) +\lambda ^{3}v^{2}\right) \varphi
_{\lambda }^{2}dxdt\geq C_{14}\lambda \dint\limits_{\Omega _{c+2\varepsilon
}}\left( \left\vert \nabla u\right\vert ^{2}+u^{2}\right) \varphi _{\lambda
}^{2}dxdt.
\end{equation*}%
Comparing this inequality with (\ref{6.12}), (\ref{6.13}) and (\ref{6.14}),
we obtain%
\begin{eqnarray}
\lambda \dint\limits_{\Omega _{c+2\varepsilon }}\left( \left\vert \nabla
u\right\vert ^{2}+u_{t}^{2}+u^{2}\right) \varphi _{\lambda }^{2}dxdt &\leq
&C_{14}\dint\limits_{Q_{T}^{\pm }}\left( \left\vert \nabla u\right\vert
^{2}+u^{2}\right) ^{2}\varphi _{\lambda }^{2}dxdt  \label{6.15} \\
&&+C_{14}\left( \left\Vert g_{0}\right\Vert _{H^{1}\left( Q_{T}^{\pm
}\right) }^{2}+\left\Vert g_{1}\right\Vert _{L_{2}\left( Q_{T}^{\pm }\right)
}^{2}+K^{2}\right) e^{3\lambda m}.  \notag
\end{eqnarray}%
Next,%
\begin{equation*}
C_{14}\dint\limits_{Q_{T}^{\pm }}\left( \left\vert \nabla u\right\vert
^{2}+u_{t}^{2}+u^{2}\right) ^{2}\varphi _{\lambda }^{2}dxdt=
\end{equation*}%
\begin{eqnarray*}
&&C_{14}\dint\limits_{\Omega _{c+2\varepsilon }}\left( \left\vert \nabla
u\right\vert ^{2}+u_{t}^{2}+u^{2}\right) ^{2}\varphi _{\lambda
}^{2}dxdt+C_{14}\dint\limits_{Q_{T}^{\pm }\diagdown \Omega _{c+2\varepsilon
}}\left( \left\vert \nabla u\right\vert ^{2}+u_{t}^{2}+u^{2}\right)
^{2}\varphi _{\lambda }^{2}dxdt \\
&\leq &C_{14}\dint\limits_{\Omega _{c+2\varepsilon }}\left( \left\vert
\nabla u\right\vert ^{2}+u_{t}^{2}+u^{2}\right) ^{2}\varphi _{\lambda
}^{2}dxdt+C_{14}e^{2\lambda \left( c+2\varepsilon \right)
}\dint\limits_{Q_{T}^{\pm }\diagdown \Omega _{c+2\varepsilon }}\left(
\left\vert \nabla u\right\vert ^{2}+u_{t}^{2}+u^{2}\right) ^{2}dxdt.
\end{eqnarray*}%
Comparing this with (\ref{6.15}) and taking $\lambda \geq \max \left(
\lambda _{0},\lambda _{1}\right) ,$ where $\lambda _{1}=C_{14}/2,$ we obtain%
\begin{eqnarray}
&&\lambda \dint\limits_{\Omega _{c+2\varepsilon }}\left( \left\vert \nabla
u\right\vert ^{2}+u_{t}^{2}+u^{2}\right) \varphi _{\lambda }^{2}dxdt\leq
C_{14}e^{2\lambda \left( c+2\varepsilon \right) }\dint\limits_{Q_{T}^{\pm
}\diagdown \Omega _{c+2\varepsilon }}\left( \left\vert \nabla u\right\vert
^{2}+u_{t}^{2}+u^{2}\right) ^{2}dxdt  \label{6.16} \\
&&+C_{14}\left( \left\Vert g_{0}\right\Vert _{H^{1}\left( Q_{T}^{\pm
}\right) }^{2}+\left\Vert g_{1}\right\Vert _{L_{2}\left( Q_{T}^{\pm }\right)
}^{2}+K^{2}\right) e^{3\lambda m}.  \notag
\end{eqnarray}%
We have $\Omega _{c+3\varepsilon }\subset \Omega _{c+2\varepsilon }$ and $%
\varphi _{\lambda }^{2}\geq \exp \left( 2\lambda \left( c+3\varepsilon
\right) \right) $ in $\Omega _{c+3\varepsilon }.$ Hence, 
\begin{equation*}
\lambda \dint\limits_{\Omega _{c+2\varepsilon }}\left( \left\vert \nabla
u\right\vert ^{2}+u_{t}^{2}+u^{2}\right) \varphi _{\lambda }^{2}dxdt\geq
\lambda e^{2\lambda \left( c+3\varepsilon \right) }\dint\limits_{\Omega
_{c+3\varepsilon }}\left( \left\vert \nabla u\right\vert
^{2}+u_{t}^{2}+u^{2}\right) dxdt.
\end{equation*}%
Substituting this inequality in (\ref{6.16}) and dividing by $\lambda
e^{2\lambda \left( c+3\varepsilon \right) },$ we obtain%
\begin{equation*}
\left\Vert u\right\Vert _{H^{1}\left( \Omega _{c+3\varepsilon }\right)
}^{2}\leq C_{14}e^{-2\lambda \varepsilon }\left\Vert u\right\Vert
_{H^{1}\left( Q_{T}^{\pm }\right) }^{2}+C_{14}\left( \left\Vert
g_{0}\right\Vert _{H^{1}\left( S_{T}^{\pm }\right) }^{2}+\left\Vert
g_{1}\right\Vert _{L_{2}\left( S_{T}^{\pm }\right) }^{2}+K^{2}\right)
e^{3\lambda m}.
\end{equation*}%
Or%
\begin{equation}
\left\Vert u\right\Vert _{H^{1}\left( \Omega _{c+3\varepsilon }\right) }\leq
C_{14}e^{-\lambda \varepsilon }\left\Vert u\right\Vert _{H^{1}\left(
Q_{T}^{\pm }\right) }+C_{14}\left( \left\Vert g_{0}\right\Vert _{H^{1}\left(
S_{T}^{\pm }\right) }+\left\Vert g_{1}\right\Vert _{L_{2}\left( S_{T}^{\pm
}\right) }+K\right) e^{\left( 3\lambda m\right) /2}.  \label{6.160}
\end{equation}

We now temporary indicate the dependence of the domain $\Omega _{c}$ on the
point $x_{0},$ i.e. $\Omega _{c}\left( x_{0}\right) .$ There exists a
sufficiently small number $c=c\left( x_{0},\left\Vert a\right\Vert
_{C^{1}\left( \overline{G}\right) }\right) >0$ and a sufficiently small
number $\varepsilon >0$ such that 
\begin{eqnarray*}
\left\{ \left\vert x-x_{0}\right\vert \leq 2\sqrt{c+3\varepsilon }\right\}
&\subset &G\text{,} \\
\left( \nabla a\left( x\right) ,x-x_{0}^{\prime }\right) &\geq &\frac{\alpha 
}{2}>0,\forall x\in \overline{G},\forall x_{0}^{\prime }\in \left\{
\left\vert x_{0}-x_{0}^{\prime }\right\vert \leq 2\sqrt{c+3\varepsilon }%
\right\} ,
\end{eqnarray*}%
see (\ref{6.2}). Choose a point $x_{0}^{\prime }$ such that $\left\vert
x_{0}-x_{0}^{\prime }\right\vert =2\sqrt{c+3\varepsilon }$. Consider a point 
$x\in G$ such that $\left\vert x-x_{0}\right\vert <\sqrt{c+3\varepsilon }.$
Hence, the point $\left( x,0\right) \notin $ $\overline{\Omega }%
_{c+3\varepsilon }\left( x_{0}\right) .$ On the other hand, by the triangle
inequality 
\begin{equation*}
\left\vert x-x_{0}^{\prime }\right\vert =\left\vert x-x_{0}-\left(
x_{0}^{\prime }-x_{0}\right) \right\vert \geq \left\vert x_{0}-x_{0}^{\prime
}\right\vert -\left\vert x-x_{0}\right\vert >2\sqrt{c+3\varepsilon }-\sqrt{%
c+3\varepsilon }=\sqrt{c+3\varepsilon }.
\end{equation*}%
Hence, $\left( x,0\right) \in \Omega _{c+3\varepsilon }\left( x_{0}^{\prime
}\right) .$ This means that $G\subset \left( \Omega _{c+3\varepsilon }\left(
x_{0}\right) \cup \Omega _{c+3\varepsilon }\left( x_{0}^{\prime }\right)
\right) .$ Hence, there exists a sufficiently small number $\sigma >0$ such
that $Q_{\sigma }^{\pm }\subset \left( \Omega _{c+3\varepsilon }\left(
x_{0}\right) \cup \Omega _{c+3\varepsilon }\left( x_{0}^{\prime }\right)
\right) :=Y.$ Clearly 
\begin{equation*}
\left\Vert u\right\Vert _{H^{1}\left( Q_{\sigma }^{\pm }\right) }\leq
\left\Vert u\right\Vert _{H^{1}\left( Y\right) }\leq \left\Vert u\right\Vert
_{H^{1}\left( \Omega _{c+3\varepsilon }\left( x_{0}\right) \right)
}+\left\Vert u\right\Vert _{H^{1}\left( \Omega _{c+3\varepsilon }\left(
x_{0}^{\prime }\right) \right) }.
\end{equation*}%
Hence, using in the left hand side of (\ref{6.160}) $\left\Vert u\right\Vert
_{H^{1}\left( \Omega _{c+3\varepsilon }\left( x_{0}\right) \right) }$ first
and $\left\Vert u\right\Vert _{H^{1}\left( \Omega _{c+3\varepsilon }\left(
x_{0}^{\prime }\right) \right) }$ and adding those two inequalities, we
obtain%
\begin{equation*}
\left\Vert u\right\Vert _{H^{1}\left( Q_{\sigma }^{\pm }\right) }\leq
C_{14}e^{-\lambda \varepsilon }\left\Vert u\right\Vert _{H^{1}\left(
Q_{T}^{\pm }\right) }+C_{14}\left( \left\Vert g_{0}\right\Vert _{H^{1}\left(
S_{T}^{\pm }\right) }+\left\Vert g_{1}\right\Vert _{L_{2}\left( S_{T}^{\pm
}\right) }+K\right) e^{\left( 3\lambda m\right) /2}.
\end{equation*}%
Hence, there exists a number $t_{0}\in \left( -\sigma ,\sigma \right) $ such
that%
\begin{equation}
\dint\limits_{\Omega }\left( \left\vert \nabla u\right\vert
^{2}+u_{t}^{2}+u^{2}\right) \left( x,t_{0}\right) dx\leq  \label{6.17}
\end{equation}%
\begin{equation*}
C_{14}\left[ e^{-2\lambda \varepsilon }\left\Vert u\right\Vert _{H^{1}\left(
Q_{T}^{\pm }\right) }^{2}+\left( \left\Vert g_{0}\right\Vert _{H^{1}\left(
Q_{T}^{\pm }\right) }^{2}+\left\Vert g_{1}\right\Vert _{L_{2}\left(
Q_{T}^{\pm }\right) }^{2}+K^{2}\right) e^{3\lambda m}\right] .
\end{equation*}

Let $y\left( x,t\right) =L_{hyp}u.$ Then $y\in L_{2}\left( Q_{T}^{\pm
}\right) $ and 
\begin{eqnarray}
L_{hyp}u &=&y\left( x,t\right) \text{ in }Q_{T}^{\pm },  \label{6.18} \\
u\left( x,t_{0}\right)  &=&u_{0}\left( x\right) ,u_{t}\left( x,t_{0}\right)
=u_{1}\left( x\right) ,  \label{6.19} \\
u &\mid &_{S_{T}^{\pm }}=g_{0}\left( x,t\right) ,\partial _{n}u\mid
_{S_{T}^{\pm }}=g_{1}\left( x,t\right) .  \label{6.20}
\end{eqnarray}%
We now refer to the classical method of energy estimates, see, e.g. chapter
4 in the book of Ladyzhenskaya \cite{Lad}. First, consider conditions (\ref%
{6.18})-(\ref{6.20}) as the initial boundary value problem in $%
Q_{t_{0},T}=\Omega \times \left( t_{0},T\right) .$ Then the method of energy
estimates gives 
\begin{eqnarray*}
&&\left\Vert u\right\Vert _{H^{1}\left( Q_{t_{0},T}\right) }^{2} \\
&\leq &C_{14}\left( \left\Vert u_{0}\right\Vert _{H^{1}\left( \Omega \right)
}^{2}+\left\Vert u_{1}\right\Vert _{L_{2}\left( \Omega \right)
}^{2}+\left\Vert g_{0}\right\Vert _{H^{1}\left( \partial \Omega \times
\left( t_{0},T\right) \right) }^{2}+\left\Vert g_{1}\right\Vert
_{H^{1}\left( \partial \Omega \times \left( t_{0},T\right) \right)
}^{2}+\left\Vert y\right\Vert _{L_{2}\left( Q_{t_{0,}T}\right) }^{2}\right) .
\end{eqnarray*}%
Next, since time can be reversed in hyperbolic PDEs, we consider conditions (%
\ref{6.18})-(\ref{6.20}) as the initial boundary value problem in $%
Q_{t_{0,}-T}=\Omega \times \left( -T,t_{0}\right) .$ Then again the energy
estimate leads to 
\begin{eqnarray*}
&&\left\Vert u\right\Vert _{H^{1}\left( Q_{t_{0},-T}\right) }^{2} \\
&\leq &C_{14}\left( \left\Vert u_{0}\right\Vert _{H^{1}\left( \Omega \right)
}^{2}+\left\Vert u_{1}\right\Vert _{L_{2}\left( \Omega \right)
}^{2}+\left\Vert g_{0}\right\Vert _{H^{1}\left( \partial \Omega \times
\left( -T,t_{0}\right) \right) }^{2}+\left\Vert g_{1}\right\Vert
_{H^{1}\left( \partial \Omega \times \left( -T,t_{0}\right) \right)
}^{2}+\left\Vert y\right\Vert _{L_{2}\left( Q_{t_{0},-T}\right) }^{2}\right)
.
\end{eqnarray*}%
Summing up the last two inequalities, we obtain%
\begin{eqnarray}
&&\left\Vert u\right\Vert _{H^{1}\left( Q_{T}^{\pm }\right) }^{2}
\label{6.22} \\
&\leq &C_{14}\left( \left\Vert u_{0}\right\Vert _{H^{1}\left( \Omega \right)
}^{2}+\left\Vert u_{1}\right\Vert _{L_{2}\left( \Omega \right)
}^{2}+\left\Vert g_{0}\right\Vert _{H^{1}\left( S_{T}^{\pm }\right)
}^{2}+\left\Vert g_{1}\right\Vert _{H^{1}\left( S_{T}^{\pm }\right)
}^{2}+\left\Vert y\right\Vert _{L_{2}\left( Q_{T}^{\pm }\right) }^{2}\right)
.  \notag
\end{eqnarray}%
By (\ref{6.5}) $\left\Vert y\right\Vert _{L_{2}\left( Q_{T}^{\pm }\right)
}^{2}\leq K^{2}.$ Hence, (\ref{6.17}) and (\ref{6.22}) lead to%
\begin{equation}
\left\Vert u\right\Vert _{H^{1}\left( Q_{T}^{\pm }\right) }^{2}\leq C_{14}
\left[ e^{-2\lambda \varepsilon }\left\Vert u\right\Vert _{H^{1}\left(
Q_{T}^{\pm }\right) }^{2}+\left( \left\Vert g_{0}\right\Vert _{H^{1}\left(
Q_{T}^{\pm }\right) }^{2}+\left\Vert g_{1}\right\Vert _{L_{2}\left(
Q_{T}^{\pm }\right) }^{2}+K^{2}\right) e^{3\lambda m}\right] .  \label{6.23}
\end{equation}%
Choose $\lambda _{2}\geq \max \left( \lambda _{0},\lambda _{1}\right) $ such
that $C_{14}e^{-2\lambda _{2}\varepsilon }<1/2.$ Then (\ref{6.23}) implies
that 
\begin{equation*}
\left\Vert u\right\Vert _{H^{1}\left( Q_{T}^{\pm }\right) }^{2}\leq
C_{14}\left( \left\Vert g_{0}\right\Vert _{H^{1}\left( Q_{T}^{\pm }\right)
}^{2}+\left\Vert g_{1}\right\Vert _{L_{2}\left( Q_{T}^{\pm }\right)
}^{2}+K^{2}\right) e^{3\lambda _{2}m}.
\end{equation*}%
This immediately leads to the target estimate (\ref{6.6}) with a new
constant $C_{14}.$

Consider now the case $a\left( x\right) \equiv 1$, $G=\left\{ \left\vert
x\right\vert <R\right\} .$ Since by Lemma 6.1 one can take $\alpha =0$ in (%
\ref{6.2}) in this case, then we choose $x_{0}=0$ and then follow the above
proof. $\square $

\subsection{Regularization}

\label{sec:6.3}

Recall that we need to find an approximate solution $u$ of the problem (\ref%
{6.3}), (\ref{6.4}). Just as above, we assume that there exists a function $%
F $ satisfying the following conditions 
\begin{equation}
F\in H^{2}\left( Q_{T}^{\pm }\right) ,F\mid _{S_{T}^{\pm }}=g_{0}\left(
x,t\right) ,\partial _{n}F\mid _{S_{T}^{\pm }}=g_{1}\left( x,t\right) .
\label{6.27}
\end{equation}%
Let $\left( ,\right) $ and $\left[ ,\right] $ be scalar products in $%
L_{2}\left( Q_{T}^{\pm }\right) $ and $H^{2}\left( Q_{T}^{\pm }\right) $
respectively and let $\left\Vert \cdot \right\Vert $ and $\left[ \cdot %
\right] $ be respective norms. We find an approximate solution of the
problem (\ref{6.3}), (\ref{6.4}) via the minimization of the following
Tikhonov functional 
\begin{eqnarray}
J_{\gamma }\left( u\right) &=&\left\Vert L_{hyp}u-f\right\Vert ^{2}+\gamma 
\left[ u-F\right] ^{2},  \label{6.28} \\
&&\text{subject to the lateral Cauchy data (\ref{6.4}).}  \label{6.29}
\end{eqnarray}%
\textbf{Theorem 6.2 }(uniqueness and existence of the minimizer)\textbf{. }%
\emph{Assume that there exists a function }$F$\emph{\ satisfying conditions (%
\ref{6.27}). Then for every }$\gamma \in \left( 0,1\right) $\emph{\ there
exists unique minimizer }$u_{\gamma }\in H^{2}\left( Q_{T}^{\pm }\right) $%
\emph{\ of the functional (\ref{6.28}), (\ref{6.29}) and the following
estimate holds }%
\begin{equation*}
\left[ u_{\gamma }\right] \leq \frac{C_{14}}{\sqrt{\gamma }}\left( \left[ F%
\right] +\left\Vert f\right\Vert _{L_{2}\left( Q_{T}^{\pm }\right) }\right) ,
\end{equation*}%
\emph{where }$C_{14}=C_{14}\left( P,a_{l},a_{u},\left\Vert \nabla
a\right\Vert _{C\left( \overline{G}\right) },M,\eta ,Q_{T}^{\pm }\right) >0$%
\emph{\ is the constant of Theorem 6.1. }

The proof of this theorem is similar with the proof of Theorem 2.4 and is,
therefore, omitted. Again, to estimate the convergence rate of minimizers,
we assume that there exists the exact solution $u^{\ast }\in H^{2}\left(
Q_{T}^{\pm }\right) $ of the problem (\ref{6.3}), (\ref{6.4}) with the exact
right hand side $f^{\ast }$ in (\ref{6.3}) and exact lateral Cauchy data $%
g_{0}^{\ast },g_{1}^{\ast }$ in (\ref{6.4}). Hence, there exists a function $%
F^{\ast }$ satisfying conditions (\ref{6.27}) with $g_{0}^{\ast
},g_{1}^{\ast }$ .

\textbf{Theorem 6.3} (convergence rate). \emph{For }$\gamma \in \left(
0,1\right) ,$\emph{\ let }$u_{\gamma }\in H^{2}\left( Q_{T}^{\pm }\right) $%
\emph{\ be the unique minimizer of the functional (\ref{6.28}), (\ref{6.29}%
), which is guaranteed by Theorem 6.2. Suppose that there exists a point }$%
x_{0}\in G$\emph{\ such that condition (\ref{6.2}) is satisfied. Let the
number }$P=P\left( x_{0},G\right) =\max_{x\in \overline{G}}\left\vert
x-x_{0}\right\vert .$\emph{\ Suppose that }$T>P/\sqrt{\eta _{0}},$\emph{\
where }$\eta _{0}=\eta _{0}\left( G,P,a_{l},a_{u},\left\Vert \nabla
a\right\Vert _{C\left( \overline{G}\right) }\right) \in \left( 0,1\right) $%
\emph{\ is the number of Lemma 6.1. Then with the constant }$%
C_{14}=C_{14}\left( P,a_{l},a_{u},\left\Vert \nabla a\right\Vert _{C\left( 
\overline{G}\right) },M,\eta _{0},Q_{T}^{\pm }\right) >0$\emph{\ of Theorem
6.1 the following estimate holds}%
\begin{equation}
\left\Vert u_{\gamma }-u^{\ast }\right\Vert _{H^{1}\left( Q_{T}^{\pm
}\right) }\leq  \label{6.31}
\end{equation}%
\begin{equation*}
C_{14}\left( \left\Vert g_{0}-g_{0}^{\ast }\right\Vert _{H^{1}\left(
S_{T}^{\pm }\right) }+\left\Vert g_{1}-g_{1}^{\ast }\right\Vert
_{L_{2}\left( S_{T}^{\pm }\right) }+\left\Vert f-f^{\ast }\right\Vert +\left[
F-F^{\ast }\right] +\sqrt{\gamma }\left[ u^{\ast }\right] \right) ,
\end{equation*}%
\emph{where }$u_{\gamma }$\emph{\ is the minimizer of the functional (\ref%
{6.28}), (\ref{6.29}), which is guaranteed by Theorem 6.2. In particular,
let }$\delta \in \left( 0,1\right) $\emph{, }$\gamma =\gamma \left( \delta
\right) =\delta ^{2}$\emph{\ and let }%
\begin{equation*}
\left\Vert g_{0}-g_{0}^{\ast }\right\Vert _{H^{1}\left( S_{T}^{\pm }\right)
},\left\Vert g_{1}-g_{1}^{\ast }\right\Vert _{L_{2}\left( S_{T}^{\pm
}\right) },\left\Vert f-f^{\ast }\right\Vert ,\left[ F-F^{\ast }\right] \leq
\delta .
\end{equation*}%
\emph{\ Then (\ref{6.31}) becomes}%
\begin{equation}
\left\Vert u_{\gamma \left( \delta \right) }-u^{\ast }\right\Vert
_{H^{1}\left( Q_{T}^{\pm }\right) }\leq C_{14}\left( 1+\left[ u^{\ast }%
\right] \right) \delta .  \label{6.32}
\end{equation}%
\emph{In the case }$a\left( x\right) \equiv 1$\emph{\ and }$G=\left\{
\left\vert x\right\vert <R\right\} $\emph{\ condition (\ref{6.2}) is not
required and estimates (\ref{6.31}), (\ref{6.32}) are valid for }$T>R.$

\textbf{Proof.} We need to prove this theorem, since, unlike all above
convergence results, we have $\delta ^{1}$ in (\ref{6.32}) instead of $%
\delta ^{\varkappa }$ with a certain $\varkappa \in \left( 0,1\right) .$
Thus, the convergence rate is stronger here than in above theorems. Denote $%
\widetilde{u}_{\gamma }=u^{\ast }-u_{\gamma },\widetilde{f}=f^{\ast }-f,%
\widetilde{F}=F^{\ast }-F.$ Then\emph{\ }we obtain \ similarly with (\ref%
{2.1160})%
\begin{eqnarray}
\left\Vert L_{hyp}\widetilde{u}_{\gamma }\right\Vert ^{2}+\gamma \left[ 
\widetilde{u}_{\gamma }\right] ^{2} &\leq &\frac{1}{2}\left\Vert L_{hyp}%
\widetilde{u}_{\gamma }\right\Vert ^{2}+\frac{1}{2}\left\Vert \widetilde{f}%
-L_{hyp}\widetilde{F}\right\Vert ^{2}  \label{6.33} \\
&&+\frac{\gamma }{2}\left[ u^{\ast }\right] ^{2}+\frac{\gamma }{2}\left[ 
\widetilde{u}_{\gamma }\right] ^{2}.  \notag
\end{eqnarray}%
Hence,%
\begin{equation}
\left\Vert L_{hyp}\widetilde{u}_{\gamma }\right\Vert ^{2}\leq \left\Vert 
\widetilde{f}-L_{hyp}\widetilde{F}\right\Vert ^{2}+\gamma \left[ u^{\ast }%
\right] ^{2},  \label{6.34}
\end{equation}%
Estimates (\ref{6.31}), (\ref{6.32}) follow immediately from (\ref{6.34})
and Theorem 6.1. The statement about the removal of the multiplier $1/\sqrt{%
\gamma }$ follows from (\ref{6.31}). The special case $a\left( x\right)
\equiv 1$\emph{, }$G=\left\{ \left\vert x\right\vert <R\right\} $ follows
from (\ref{6.34}) and Theorem 6.1. $\square $

\section{Thermoacoustic tomography}

\label{sec:7}

In this section we show how results of section 6 can be applied to the
problem of the reconstruction of one of initial conditions of a hyperbolic
PDE from boundary measurements. This problem is called nowadays
\textquotedblleft the problem of thermoacoustic tomography". Although
results of this section actually follow from the earlier work of Klibanov
and Malinsky \cite{KM} (1991), this problem did not have that name at that
time. More details can be found in the paper of the author \cite{Kltherm}.
Numerical studies by the method of this section were performed in \cite%
{ClK,KKKN}.

\subsection{Statement of the inverse problem}

\label{sec:7.1}

We assume below that the domain $G$ is a ball of the radius $R$, $G=\left\{
x\in \mathbb{R}^{n}:\left\vert x\right\vert <R\right\} $ and we keep
notations (\ref{6.0}). Although a more general domain can be considered
along the same lines, we are not doing so for brevity. Denote $D_{T}^{n+1}=%
\mathbb{R}^{n}\times \left( 0,T\right) .$ Let the function $u\in H^{2}\left(
D_{T}^{n+1}\right) $ be the solution of the following Cauchy problem%
\begin{equation}
\widehat{L}_{hyp}\left( u\right) =a\left( x\right) u_{tt}-\Delta
u-\dsum\limits_{j=1}^{n}b_{j}\left( x\right) u_{x_{j}}-b_{0}\left( x\right)
u=0\text{ in }D_{T}^{n+1},  \label{6.35}
\end{equation}%
\begin{equation}
u\left( x,0\right) =f\left( x\right) ,u\left( x,0\right) =0.  \label{6.36}
\end{equation}%
We impose the following conditions on coefficients of equation (\ref{6.35})%
\begin{eqnarray}
a &\in &C^{1}\left( \mathbb{R}^{n}\right) ,a\left( x\right) =1\text{ and for 
}x\in \mathbb{R}^{n}\diagdown G,  \label{6.37} \\
b_{j} &\in &C\left( \mathbb{R}^{n}\right) \text{ and }b_{j}\left( x\right)
=0,j=0,...,n\text{ for }x\in \mathbb{R}^{n}\diagdown G,  \label{6.38} \\
B &=&\max_{j}\left\Vert b_{j}\right\Vert _{C\left( \overline{G}\right) }.
\label{6.39}
\end{eqnarray}%
In addition, we impose conditions (\ref{6.1}) on the coefficient $a\left(
x\right) .$ Finally, we assume that 
\begin{equation}
f\in H^{3}\left( \mathbb{R}^{n}\right) \text{ and }f\left( x\right) =0\text{
for }x\in \mathbb{R}^{n}\diagdown G.  \label{6.40}
\end{equation}%
Corollary 4.1 of \S 4 of Chapter 4 of the book of Ladyzhenskaya \cite{Lad}
guarantees that there exists unique solution $u\in H^{3}\left(
D_{T}^{n+1}\right) $ of the Cauchy problem (\ref{6.35}), (\ref{6.36}), as
long as conditions (\ref{6.37}), (\ref{6.40}) are satisfied.

\textbf{Inverse Problem 1}. Find the initial condition $f\left( x\right) $
assuming that the function $p\left( x,t\right) $ is given, where%
\begin{equation}
u\mid _{S_{T}}=p\left( x,t\right) .  \label{6.41}
\end{equation}

Hence, $p\in H^{2}\left( S_{T}\right) $ in the case of exact data. In the
case of real measurements, the function $p$ is given with a noise. However,
it can be smoothed out by a number of well known procedures, so that the
resulting function belongs to $H^{2}\left( S_{T}\right) .$ This is an
inverse problem of finding the initial condition from boundary
measurements.\ In the case $b_{j}\left( x\right) \equiv 0,j=0,...,n$ this
problem is called sometimes \textquotedblleft the problem of thermoacoustic
tomography". Following the technique of section 6, we need to figure out the
Neumann boundary condition at $S_{T}$ and to estimate it somehow via the
function $p\left( x,t\right) .$ Instead, we will consider a ball, which is
both concentric with $G$ and larger than $G$, find both Dirichlet and
Neumann boundary conditions on its boundary and estimate them through $%
p\left( x,t\right) .$\ 

Let $\sigma >0$ be a number. Denote 
\begin{equation*}
G^{\sigma }=\left\{ x\in \mathbb{R}^{n}:\left\vert x\right\vert <R+\sigma
\right\} ,Q_{T}^{\sigma }=G^{\sigma }\times \left( 0,T\right) ,S_{T}^{\sigma
}=\partial G^{\sigma }\times \left( 0,T\right) ,
\end{equation*}%
\begin{equation}
g_{0}^{\sigma }\left( x,t\right) =u\mid _{S_{T}^{\sigma }},g_{1}^{\sigma
}\left( x,t\right) =\partial _{n}u\mid _{S_{T}^{\sigma }}.  \label{6.42}
\end{equation}%
Similarly with the above, below in this section $C_{15}=C_{15}\left(
a_{l},a_{u},B,R,\sigma ,T\right) >0$ denotes different positive constants
depending on listed parameters.

\textbf{Lemma 7.1}. \emph{Let conditions (\ref{6.1}), (\ref{6.35})-(\ref%
{6.42}) be satisfied. Also, let the function }$p\in H^{2}\left( S_{T}\right)
.$\emph{\ Then there exists a number }$\overline{\sigma }\in \left( 0,\sigma
\right) $ \emph{and a number }

$C_{15}=C_{15}\left( a_{l},a_{u},B,R,\sigma ,T\right) >0$\emph{\ such that }%
\begin{equation}
\left\Vert g_{0}^{\overline{\sigma }}\right\Vert _{H^{1}\left( S_{T}^{%
\overline{\sigma }}\right) }\leq C_{15}\left\Vert p\right\Vert _{H^{2}\left(
S_{T}\right) },\left\Vert g_{1}^{\overline{\sigma }}\right\Vert
_{L_{2}\left( S_{T}^{\overline{\sigma }}\right) }\leq C_{15}\left\Vert
p\right\Vert _{H^{2}\left( S_{T}\right) }.  \label{6.43}
\end{equation}

\textbf{Proof}. For $r>0,$ consider the function $\phi _{1}\left( r\right) $
such that 
\begin{equation*}
\phi _{1}\left( r\right) \in C^{2}\left[ 0,\infty \right) ,\phi _{1}\left(
r\right) =\left\{ 
\begin{array}{c}
1,r\in \left[ 0,R+2\sigma \right] , \\ 
\in \left( 0,1\right) ,r\in \left( R+2\sigma ,R+3\sigma \right) , \\ 
0,r\geq R+3\sigma .%
\end{array}%
\right.
\end{equation*}%
For $\left\vert x\right\vert \geq R$ consider functions $q\left( x,t\right) $
and $v\left( x,t\right) $ defined as%
\begin{equation}
q\left( x,t\right) =\phi _{1}\left( \left\vert x\right\vert \right) p\left(
x,t\right) ,v\left( x,t\right) =u\left( x,t\right) -q\left( x,t\right) .
\label{6.44}
\end{equation}%
Then (\ref{6.35})-(\ref{6.41}) and (\ref{6.44}) imply that the function is
the solution of the following initial boundary value problem outside of the
ball $G$%
\begin{eqnarray}
\widehat{L}_{hyp}\left( v\right) &=&-\widehat{L}_{hyp}\left( q\right)
,\left( x,t\right) \in \left( \mathbb{R}^{n}\diagdown G\right) \times \left(
0,T\right) ,  \notag \\
v\left( x,0\right) &=&v_{t}\left( x,0\right) =0,x\in \mathbb{R}^{n}\diagdown
G,  \label{6.45} \\
v &\mid &_{S_{T}}=0.  \notag
\end{eqnarray}%
Also, $v\in H^{2}\left( \left( \mathbb{R}^{n}\diagdown G\right) \times
\left( 0,T\right) \right) .$ Consider a ball $G^{\prime }=\left\{ x\in 
\mathbb{R}^{n}:\left\vert x\right\vert <R^{\prime }\right\} $ where the
number $R^{\prime }=R^{\prime }\left( a_{l},a_{u},B,\sigma ,R,T\right)
>R+3\sigma $ is so large that $u\left( x,t\right) =0$ for $\left( x,t\right)
\in \left\{ \left\vert x\right\vert >R^{\prime }/2\right\} \times \left(
0,T\right) .$ Such a number $R^{\prime }$ exists due to the finite speed of
propagation of solutions of hyperbolic equations, see \S 2 of chapter 4 of
the book of Ladyzhenskaya \cite{Lad}. Applying the method of energy
estimates \cite{Lad} to the problem (\ref{6.45}) in the domain $%
P_{T}=\left\{ \left( x,t\right) :R<\left\vert x\right\vert <R^{\prime },t\in
\left( 0,T\right) \right\} $, taking into account zero Dirichlet boundary
conditions for the function $v$ at $\left\{ \left\vert x\right\vert
=R\right\} \cup \left\{ \left\vert x\right\vert =R^{\prime }\right\} $ and
then taking into account (\ref{6.44}), we obtain%
\begin{equation}
\left\Vert u\right\Vert _{H^{1}\left( \left( G^{\sigma }\diagdown G\right)
\times \left( 0,T\right) \right) }\leq \left\Vert u\right\Vert _{H^{1}\left(
P_{T}\right) }\leq C_{15}\left\Vert p\right\Vert _{H^{2}\left( S_{T}\right)
}.  \label{6.46}
\end{equation}%
Since there exists a number $\overline{\sigma }\in \left( 0,\sigma \right) $
such that 
\begin{equation*}
\left\Vert u\right\Vert _{H^{1}\left( S_{T}^{\overline{\sigma }}\right)
},\left\Vert \partial _{n}u\right\Vert _{L_{2}\left( S_{T}^{\overline{\sigma 
}}\right) }\leq \frac{1}{\sigma }\left\Vert u\right\Vert _{H^{1}\left(
\left( G^{\sigma }\diagdown G\right) \times \left( 0,T\right) \right) },
\end{equation*}%
then (\ref{6.46}) completes the proof. $\square $

\subsection{Lipschitz stability}

\label{sec:7.2}

\textbf{Theorem 7.1}. \emph{Let }$\sigma >0$\emph{\ be a number.} \emph{Let
conditions (\ref{6.1}), (\ref{6.2}), (\ref{6.35})-(\ref{6.41}) be satisfied.
Let the function }$u\in H^{3}\left( D_{T}^{n+1}\right) $\emph{\ be the
solution of the problem (\ref{6.35}), (\ref{6.36}) and let (\ref{6.41}) be
valid. Let the number }$P=P\left( x_{0},R\right) =\max_{x\in \overline{G}%
}\left\vert x-x_{0}\right\vert .$\emph{\ Then there exists a number }$\eta
_{0}=\eta _{0}\left( R,x_{0},a_{l},a_{u},\left\Vert \nabla a\right\Vert
_{C\left( \overline{G}\right) }\right) \in \left( 0,1\right) $\emph{\ such
that if }$T>P/\sqrt{\eta _{0}},$\emph{\ then there exists a number }$%
C_{15}=C_{15}\left( a_{l},a_{u},B,R,\sigma ,\eta _{0},T\right) >0$\emph{\
such that the following Lipschitz stability estimate holds }$\left\Vert
f\right\Vert _{L_{2}\left( G\right) }\leq C_{15}\left\Vert p\right\Vert
_{H^{2}\left( S_{T}\right) }.$

\textbf{Proof.} Let $\overline{\sigma }\in \left( 0,\sigma \right) $ be the
number of Lemma 7.1. To apply Theorem 6.1, we replace in it first $G$ with $%
G^{\overline{\sigma }}$. We notice that when we integrate the pointwise
Carleman estimate (\ref{2.49}) in the proof of Theorem 6.1 over the domain $%
\Omega _{c}\cap \left\{ t>0\right\} $ and use the Gauss' formula, the
boundary integral over $\Omega _{c}\cap \left\{ t=0\right\} $ equals zero
because of (\ref{2.52}) and also because $u_{t}\left( x,0\right) =0$ by (\ref%
{6.36}).The rest of the proof is identical to the rest of the proof of
Theorem 6.1. Hence, using Lemma 7.1, we obtain $\left\Vert u\right\Vert
_{H^{1}\left( G^{\overline{\sigma }}\times \left( 0,T\right) \right) }\leq
C_{15}\left\Vert p\right\Vert _{H^{2}\left( S_{T}\right) }.$ Since $G\subset
G^{\overline{\sigma }},$ then by the trace theorem $\left\Vert f\right\Vert
_{L_{2}\left( G\right) }\leq C_{15}\left\Vert p\right\Vert _{H^{2}\left(
S_{T}\right) }.$ $\ \square $

The fact that this theorem depends on a number $\sigma >0$ is a minor issue
in this context. The author believes that this dependence can be eliminated.
Since this likely would require an extensive effort, the author is not doing
this here.

\subsection{Regularization}

\label{sec:7.3}

We would need to use now $G^{\overline{\sigma }}.$ However, for brevity we
replace here $G^{\overline{\sigma }}$ with $G.$ This also makes sense from
the computational point of view \cite{ClK,KKKN}. Indeed, in order to solve
Inverse Problem 1 in practical computations, one should first solve the
initial boundary value problem in $\left( \mathbb{R}^{n}\diagdown G\right)
\times \left( 0,T\right) ,$ 
\begin{eqnarray*}
\widehat{L}_{hyp}\left( u\right)  &=&0,\left( x,t\right) \in \left( \mathbb{R%
}^{n}\diagdown G\right) \times \left( 0,T\right) , \\
u\left( x,0\right)  &=&u_{t}\left( x,0\right) =0,x\in \mathbb{R}%
^{n}\diagdown G, \\
u &\mid &_{S_{T}}=p\left( x,t\right) .
\end{eqnarray*}%
This way one finds the function $\overline{p}\left( x,t\right) =\partial
_{n}u\mid _{S_{T}}.$ Hence, we construct now a numerical method for Inverse
Problem 2. Let 
\begin{eqnarray}
\widehat{L}_{hyp}\left( u\right)  &=&0\text{ in }Q_{T},u\in H^{2}\left(
Q_{T}\right) ,  \label{100} \\
u &\mid &_{S_{T}}=p\left( x,t\right) ,\partial _{n}u\mid _{S_{T}}=\overline{p%
}\left( x,t\right) ,  \label{101} \\
u_{t}\left( x,0\right)  &=&0.  \label{102}
\end{eqnarray}

\textbf{Inverse\ Problem 2}. Suppose that functions $p,\overline{p}$ in (\ref%
{101}) are given. Determine the function $f\left( x\right) =u\left(
x,0\right) $ for $x\in G$ from conditions (\ref{100})-(\ref{102}).

The difference between the problem (\ref{100})-(\ref{102}) and the problem (%
\ref{6.3}), (\ref{6.4}) is that now we require one initial condition (\ref%
{102}). We replace conditions (\ref{6.37})-(\ref{6.39}) with 
\begin{equation}
a\in C^{1}\left( \overline{G}\right) ,b_{j}\in C\left( \overline{G}\right)
,B=\max_{j}\left\Vert b_{j}\right\Vert _{C\left( \overline{G}\right) },\text{
}j=0,...,n\text{ .}  \label{6.47}
\end{equation}%
Suppose that there exists a function $F\in H^{2}\left( Q_{T}\right) $ such
that%
\begin{equation}
F\mid _{S_{T}}=p\left( x,t\right) ,\partial _{n}F\mid _{S_{T}}=\overline{p}%
\left( x,t\right) ,F_{t}\left( x,0\right) =0.  \label{6.48}
\end{equation}%
Similarly with (\ref{6.28}), (\ref{6.29}) introduce the following Tikhonov
functional with the regularization parameter $\gamma \in \left( 0,1\right) $%
\begin{eqnarray}
\overline{J}_{\gamma }\left( u\right)  &=&\left\Vert \widehat{L}_{hyp}\left(
u\right) \right\Vert _{L_{2}\left( Q_{T}\right) }^{2}+\gamma \left\Vert
u-F\right\Vert _{H^{2}\left( Q_{T}\right) }^{2},  \label{6.49} \\
&&\text{subject to conditions (\ref{101}), (\ref{102}).}  \label{6.50}
\end{eqnarray}%
Theorem 7.2 is a full analog of Theorem 2.4. Therefore, we omit its proof.

\textbf{Theorem 7.2 }(uniqueness and existence of the minimizer)\textbf{. }%
\emph{Assume that there exists a function }$F$\emph{\ satisfying conditions (%
\ref{6.48}). Then for every }$\gamma \in \left( 0,1\right) $\emph{\ there
exists unique minimizer }$u_{\gamma }\in H^{2}\left( Q_{T}\right) $\emph{\
of the functional (\ref{6.49}), (\ref{6.50}) and the following estimate
holds with the constant }$C_{15}>0$ depending on the same parameters as in
Theorem 6.4\emph{\ }%
\begin{equation*}
\left\Vert u_{\gamma }\right\Vert _{H^{2}\left( Q_{T}\right) }\leq \frac{%
C_{15}}{\sqrt{\gamma }}\left\Vert F\right\Vert _{H^{2}\left( Q_{T}\right) }.
\end{equation*}

To prove convergence of minimizers of the functional (\ref{6.49}), (\ref%
{6.50}), we again introduce the exact solution $u^{\ast }\in H^{2}\left(
Q_{T}\right) $ of the problem (\ref{100})-(\ref{102}) with exact boundary
data $p^{\ast },\overline{p}^{\ast }.$ Let $f^{\ast }\left( x\right)
=u^{\ast }\left( x,0\right) $. Then there exists a function $F^{\ast }\in
H^{2}\left( Q_{T}\right) $ satisfying conditions (\ref{6.48}) with $p^{\ast
},\overline{p}^{\ast }$ in them. We omit the proof of Theorem 7.3, since it
follows immediately from Theorems 7.1 and 7.2 in the same way as Theorem 6.3
follows from Theorems 6.1 and 6.2.

\textbf{Theorem 7.3} (convergence rate). \emph{Let conditions (\ref{6.1}), (%
\ref{6.47}) be satisfied and assume the existence of such am point }$%
x_{0}\in G$\emph{\ that condition (\ref{6.2}) is satisfied as well. For }$%
\gamma \in \left( 0,1\right) ,$\emph{\ let }$u_{\gamma }\in H^{2}\left(
Q_{T}\right) $\emph{\ be the unique minimizer of the functional (\ref{6.49}%
), (\ref{6.50}), which is guaranteed by Theorem 7.2. Denote }$P=P\left(
x_{0},G\right) =\max_{x\in \overline{G}}\left\vert x-x_{0}\right\vert .$%
\emph{\ Suppose that }$T>P/\sqrt{\eta _{0}},$\emph{\ where }$\eta _{0}=\eta
_{0}\left( G,P,a_{l},a_{u},\left\Vert \nabla a\right\Vert _{C\left( 
\overline{G}\right) }\right) \in \left( 0,1\right) $\emph{\ is the number of
Lemma 6.1. Then with the constant }$C_{16}=C_{16}\left(
G,P,a_{l},a_{u},\left\Vert \nabla a\right\Vert _{C\left( \overline{G}\right)
},B,\eta _{0},T\right) >0$\emph{\ the following estimates hold}%
\begin{equation}
\left\Vert u_{\gamma }\left( x,0\right) -f^{\ast }\left( x\right)
\right\Vert _{L_{2}\left( G\right) },\left\Vert u_{\gamma }-u^{\ast
}\right\Vert _{H^{1}\left( Q_{T}\right) }\leq  \label{6.51}
\end{equation}%
\begin{equation*}
C_{16}\left( \left\Vert p-p^{\ast }\right\Vert _{H^{1}\left( S_{T}\right)
}+\left\Vert \overline{p}-\overline{p}^{\ast }\right\Vert _{L_{2}\left(
S_{T}\right) }+\left\Vert F-F^{\ast }\right\Vert _{H^{2}\left( Q_{T}\right)
}+\sqrt{\gamma }\left\Vert u^{\ast }\right\Vert _{H^{2}\left( Q_{T}\right)
}\right) ,
\end{equation*}%
\emph{where }$u_{\gamma }$\emph{\ is the minimizer of the functional (\ref%
{6.48}), (\ref{6.49}), which is guaranteed by Theorem 7.2. In particular,
let }$\delta \in \left( 0,1\right) $\emph{, }$\gamma =\gamma \left( \delta
\right) =\delta ^{2}$\emph{\ and let }%
\begin{equation*}
\left\Vert p-p^{\ast }\right\Vert _{H^{1}\left( S_{T}\right) },\left\Vert 
\overline{p}-\overline{p}^{\ast }\right\Vert _{L_{2}\left( S_{T}\right)
},\left\Vert F-F^{\ast }\right\Vert _{H^{2}\left( Q_{T}\right) }\leq \delta .
\end{equation*}%
\emph{\ Then (\ref{6.51}) becomes}%
\begin{equation*}
\left\Vert u_{\gamma }\left( x,0\right) -f^{\ast }\left( x\right)
\right\Vert _{L_{2}\left( G\right) },\left\Vert u_{\gamma }-u^{\ast
}\right\Vert _{H^{1}\left( Q_{T}\right) }\leq C_{16}\left( 1+\left\Vert
u^{\ast }\right\Vert _{H^{2}\left( Q_{T}\right) }\right) \delta .
\end{equation*}%
\emph{In the case }$a\left( x\right) \equiv 1$\emph{\ condition (\ref{6.2})
is not necessary and one can choose }$\eta _{0}=1$\emph{\ and }$T>R.$

\section{Published Results}

\label{sec:8}

In this section we overview main published results about the topic of the
current paper: Tikhonov functionals for ill-posed Cauchy problems for PDEs,
which\ are generated by differential operators of those PDEs, under the
condition that the corresponding PDO admits a Carleman estimate. We consider
both linear and nonlinear problems. In addition to works on this topic of
the author with coauthors cited in Introduction, a number of quite elegant
results were obtained by Bourgeois and Dard\'{e}. They have done this for
the Cauchy problem for the Laplace equation and related problems. This
effort was initiated by Bourgeois in 2005 \cite{B1}. Papers \cite%
{B1,B3,B5,B6,B7,D2} of these authors contain quite good results of numerical
experiments. These results are obtained using the FEM. Regular $C^{0}$
finite elements were used in \cite{B1,B9,D2}. In \cite{B3,B5,B6,B7,B8}
non-conforming finite elements were used. Papers \cite%
{B1,B2,B3,B4,B6,B7,B8,D2} work with the variational formulation of the
Tikhonov functional for the Cauchy problem for the Laplace equation. That
functional is generated by the Laplace operator $\Delta .$ The paper \cite%
{B5} works with the variational formulation of the Tikhonov functional
generated by the operator $P=\Delta +k,k=const.\in \mathbb{R}$ for the
Cauchy problem for the equation $Pu=0.$

\subsection{Linear problems}

\label{sec:8.1}

All above theorems rely on $H^{2}$ spaces. It was observed in \cite{B1} that
these spaces would lead either to $C^{1}$ finite elements or to finite
differences. However, for rather complicated domains finite elements are
better applicable than finite differences. On the other hand, since $C^{0}$
finite elements are the most popular ones. Thus, the main idea of the paper 
\cite{B1} is to present a mixed formulation of the QRM, which would enable
one to work with standard $C^{0}$ finite elements. Two regularization
parameters were used in \cite{B1}.

It is clear from sections 2-7 that a stability estimate for an ill-posed
Cauchy problem implies a similar estimate for the convergence rate of
minimizers of the Tikhonov functional generated by the corresponding PDO. In
theorems of sections 2-4 rates of convergence of minimizers of Tikhonov
functionals are given only in certain subdomains of original domains, and
these rates are of the H\"{o}lder type.\ On the other hand, Theorem 5.4
provides the H\"{o}lder type convergence rate for a subdomain $%
Q_{T-\varepsilon }$ of the domain $Q_{T}$ and the logarithmic type
convergence rate for the entire domain $Q_{T}.$ Thus, one can anticipate
that only the logarithmic type convergence rate can be obtained in elliptic
and parabolic cases if considering the whole domain. The logarithmic
convergence rate in the whole domain was obtained in \cite{B4} for the case
of the Cauchy problem for the Laplace equation. To get that convergence
rate, a result of Phung \cite{P} was used. The result of \cite{P}, in turn
is based on Carleman estimates. While the result of \cite{P} is valid for
domains with the $C^{\infty }$ boundary, in \cite{B4} it was generalized for
the case of domains of the class $C^{1,1}.$ The result of \cite{B4} was
extended by Bourgeois and Dard\'{e} in \cite{B5} to the case of Lipschitz
domains and for the above operator $P=\Delta +k$.

As to the issue of the $H^{2}$ smoothness of solutions, Bourgeois and Dard%
\'{e} have made a point in \cite{B6}\ that noisy data are not smooth. Thus,
they have addressed in \cite{B6} the problem of working with non-smooth
noisy data by QRM for the case of the Cauchy problem for the Laplace
equation. They have used a duality-based approach. Dard\'{e}, Hannukainen
and Hyv\"{o}nen \cite{D2} have further extended the idea of \cite{B1}, for
the case of the Cauchy problem for an elliptic PDE, in order to work with
the standard $C^{0}$ finite elements. Furthermore, they have proven, for the
first time, a quite intriguing result about the monotonic convergence of
regularized solutions of the QRM. In other words, they have proven that a
certain norm of a certain difference between the regularized solution and
the exact solution strictly monotonically decreases when the regularization
parameter $\gamma $ decreases. Another salient feature of \cite{D2} is that
the first 3-d computations of QRM are presented there, whereas only 2-d
cases were considered numerically prior to \cite{D2}. Computations in \cite%
{B6,D2}, were performed for both cases: the \textquotedblleft pure" Cauchy
problem and the inverse obstacle problem (see subsection 7.2 for the latter).

Klibanov \cite{Klib2006} has studied the problem of determining the initial
condition of a general parabolic PDE of the second order from lateral Cauchy
data. This is of course the problem in the whole time cylinder rather than
in its part. Thus, a logarithmic stability estimate should be anticipated.
This estimate was obtained in \cite{Klib2006} using Carleman estimates.\ In
this case two Carleman estimates were combined: one for the lateral Cauchy
data and the second one for the case of reversed time. In other words,
analogs of Lemmata 4.1 and 5.3 respectively were combined. Next, the
Tikhonov functional generated by that parabolic operator was constructed and
the logarithmic convergence rate of minimizers was established.

\subsection{Nonlinear problems I: Inverse obstacle problems}

\label{sec:8.2}

Bourgeois and Dard\'{e} were the first ones who have applied the QRM to the
inverse obstacle problems. Note that these problems are nonlinear, unlike
all problems considered above. In a quite elegant work \cite{B7} they have
proposed a new iterative procedure of the predictor-corrector type. On the
predictor step they solve the Cauchy problem for the Laplace equation in a
2-d domain, which is located between the boundary of the original `large'
domain and the boundary of a first guess for the unknown obstacle. On the
corrector step they use that QRM solution to update the boundary of the
unknown obstacle via a version of the level set method. In \cite{B8} and 
\cite{B9} they have extended the idea of \cite{B7} to the much more
complicated cases of identification of defects for the elastic-plastic
constitutive law and the inverse obstacle problem for the Stokes system
respectively. Extending the idea of \cite{B1}, two mixed formulations were
used in \cite{B9} for the QRM in order to work with the standard Lagrange
finite elements.

\subsection{Nonlinear problems II: Coefficient Inverse Problems (CIPs)}

\label{sec:8.3}

A CIP is about the reconstruction of a coefficient of a PDE from boundary
measurements. Both the most important and the most challenging question in a
numerical treatment of a CIP is: \emph{Is it possible to have a rigorous
guarantee of obtaining at least one point in a small neighborhood of the
exact solution, provided that this neighborhood is unknown in advance?} The
author calls a numerical method addressing this question \emph{globally
convergent}. There are currently three types of globally convergent
numerical methods which are not only developed analytically but tested
numerically as well. First two types work for the case of a single
measurement event and the third type works for the case of multiple
measurements. First two types of methods are rooted to the original idea of 
\cite{BukhK}, since both eliminate the unknown coefficient from the original
PDE via the differentiation with respect to a certain parameter from which
this coefficient does not depend. Finally, the third type of globally
convergent numerical methods is the method of Kabanikhin and Shishlenin \cite%
{Kab,Kab1,Kab2}, which is based on a multidimensional analog of the
Gel'fand-Levitan-Krein equation.

The reason of the importance of the topic of global convergence is that
conventional least squares functionals for CIPs suffer from the phenomenon
of multiple local minima and ravines.\ Therefore, any optimization technique
for such a functional is a \emph{locally convergent} method, such as, e.g.
gradient and Newton methods.\ In other words, it has a rigorous guarantee of
convergence only if its starting point is located in a sufficiently small
neighborhood of the correct solution. Section 5.8.4 of \cite{BK}, as well as
publications \cite{KNT,Liu} contain numerical examples showing that locally
convergent methods do not converge to the correct solutions even if starting
point from the background medium, whereas the globally convergent methods
converge. Those examples are for experimental data in \cite{BK,Liu} and for
computationally simulated data in \cite{KNT}.

The first type of globally convergent numerical techniques is the
Beilina-Klibanov method for CIPs for the following hyperbolic equation 
\begin{equation}
c\left( x\right) u_{tt}=\Delta u,u\left( x,0\right) =0,x\in \mathbb{R}%
^{3},t>0.  \label{7.1}
\end{equation}%
This method has been developed since the work \cite{BKSISC}. The root in 
\cite{BukhK} is due to the fact that the unknown coefficient $c\left(
x\right) $ is \textquotedblleft eliminated" from the equation obtained by
the Laplace transform of (\ref{7.1}) via the differentiation with respect to
the parameter $s>0$ of this transform and obtaining a nonlinear integral
differential equation this way. Global convergence results can be found in 
\cite{BK,BKJIIP}. In addition to the convergence theory, this method is
currently completely verified on experimental data, see, e.g. \cite%
{BK,BTKF,BKadap,KBKSNF,TBKF}.

One of procedures of this method is the iterative solution of the boundary
value problems for certain elliptic PDEs. Boundary conditions for these
problems are generated by the boundary data for the CIP. However, in the
case of backscattering data, measurements of the function $u$ are performed
only on the backscattering side $\Gamma $ of the boundary $\partial \Omega $
of the domain of interest $\Omega .$ This means that those boundary
conditions are known only on $\Gamma $ in this case. One way to handle this
is a heuristic one: we complement the Dirichlet boundary condition on $%
\Gamma $ by such a Dirichlet boundary condition on $\partial \Omega
\diagdown \Gamma $ which is taken from the solution of the forward problem (%
\ref{7.1}) for $c\left( x\right) \equiv 1:$ we assume that the coefficient $%
c\left( x\right) =1$ for $x\in \mathbb{R}^{3}\diagdown \Omega $ and $c\left(
x\right) $ is unknown in $\Omega .$ This way has proved to work well both
for backscattering synthetic data \cite{BKJIIP} and for backscattering
experimental data \cite{BTKF,BKadap,TBKF}. Furthermore, Chow and Zou \cite%
{Chow} have shown numerically on synthetic data that the correct boundary
condition on $\partial \Omega \diagdown \Gamma $ results in basically the
same image as the complemented one as above.

The second way to work with backscattering data by the method of \cite{BK}
is to solve Cauchy problems for those elliptic equations with Dirichlet and
Neumann boundary conditions on $\Gamma .$ To solve such a problem, the
Tikhonov functional, which is generated by the corresponding elliptic
operator, should be minimized, as in section 3. This was done in chapter 6
of \cite{BK} for synthetic data. Furthermore, this way has proved to be
especially effective for such experimental data, where only one experimental
curve per each target was measured, which led to a 1-d CIP \cite{KBKSNF}.
Global convergence theorems for this case were proved in \cite{BK,KBKSNF}
using Carleman estimates. Since the theory was not the main goal of \cite%
{KBKSNF}, the proof of the global convergence was incomplete in this work.
That proof was later completed by Ozbilge \cite{O}.

The second type of globally convergent numerical methods was initiated by
the author for a CIP for equation (\ref{7.1}) \cite{Klib97} as well as for
the similar parabolic equation \cite{Kpar}. Next, this idea was modified by
the author and Timonov \cite{KT}. Recently there is a renewed interest in
this topic for the case of CIPs for equation (\ref{7.1}), see Beilina and
Klibanov \cite{BKconv} and Klibanov and Th\`{a}nh \cite{KNT}.

Loosely speaking, the first step of works \cite{BKconv,Klib97,Kpar,KNT} is
the same as in the method of \cite{BukhK,Bukh,K92,Ksurvey}: the unknown
coefficient $c\left( x\right) $ is \textquotedblleft eliminated" from the
original PDE via the differentiation either with respect to $t$, if working
in the time domain as in \cite{BKconv,Klib97,Kpar}, or with respect to the
parameter $s$ of the Laplace transform, if working in the \textquotedblleft
Laplace transform domain" \cite{KNT}. This way a nonlinear integral
differential equation is obtained with respect to a function $w$, which is
associated with the function $u\left( x,t\right) $ in (\ref{7.1}). Since
both Dirichlet and Neumann boundary conditions are available for $w$, then
this can be considered as an ill-posed nonlinear Cauchy problem. The second
step is that to solve this problem, a weighted Tikhonov functional $%
J_{\lambda }\left( w\right) $ is constructed. Similarly with the topic of
this paper, this functional is generated by the nonlinear integral
differential operator of that equation for $w$. The key element of $%
J_{\lambda }\left( w\right) $ is the presence of the Carleman Weight
Function (CWF) with the large parameter $\lambda >>1.$ The CWF is the one
which is involved in the Carleman estimate for a certain associated PDO. The
main result then claims that, given a finite set $\Phi \left( d\right) $ of
an arbitrary diameter $d$ in a certain Hilbert space, one can choose such a
value $\lambda _{0}\left( d\right) >>1$ that for all $\lambda \geq \lambda
_{0}\left( d\right) $ the functional $J_{\lambda }\left( w\right) $ is
strictly convex on the set $\Phi \left( d\right) .$ In accordance with the
Tikhonov concept for ill-posed problems \cite{BK,Tikh}, the set $\Phi \left(
d\right) $ can be considered as the set of admissible parameters and is
usually known \emph{a priori}. The strict convexity of $J_{\lambda }\left(
w\right) $, in turn implies that the gradient method of the minimization of $%
J_{\lambda }\left( w\right) $ converges to the unique minimizer of $%
J_{\lambda }\left( w\right) $ on the set $\Phi \left( d\right) $ if starting
from an arbitrary point of this set \cite{BKconv,KNT}. Since restrictions on
the diameter $d$ of the set $\Phi \left( d\right) $ are not imposed, then
this is the \emph{global convergence}. Numerical reconstruction results for
computationally simulated data in the 1d case obtained in \cite{KNT}
indicate that this type of methods is promising.

Recently Baudouin, de Buhan and Ervedoza \cite{Baud} have proposed a similar
idea for a CIP for the hyperbolic equation $u_{tt}=\Delta u+q\left( x\right)
u,x\in \Omega \mathbb{\subset R}^{n},t>0$ with the unknown potential $%
q\left( x\right) $ and with the data generated by a single measurement
event. Here $\Omega $ is a bounded domain. In \cite{Baud} a sequence of
weighted Tikhonov functionals generated by a sequence of linear hyperbolic
operators is constructed. As weights, CWFs for the operator $\partial
_{t}^{2}-\Delta $ are used. Similarly with the above, it is assumed that the
upper bound for the function $q$ is known \emph{a priori}, i.e. the number $%
B $ is known, where $\left\Vert q\right\Vert _{L_{\infty }\left( \Omega
\right) }\leq B$. As a result, the global convergence to the correct
solution $q_{correct}$ of some functions associated with minimizers of these
functionals is proven in \cite{Baud}.

\begin{center}
\textbf{Acknowledgments}
\end{center}

The author is grateful to Professors Anatoly B. Bakushinsky and Laurent
Bourgeois for their critical remarks, which have helped to improve the
quality of the presentation in this paper.\ The author is also grateful to
Professor Anatoly G. Yagola for prompting him to work on this survey.

\end{document}